\documentclass[useAMS,usenatbib,usegraphicx]{mn2e}

\usepackage[T1]{fontenc}
\usepackage{aecompl}

\title[Accurate photometry for the XDF with the CHEFs]{Accurate PSF-matched photometry and photometric redshifts for the Extreme Deep Field with the Chebyshev-Fourier functions}
\author[Y. Jim\'enez-Teja et al.]{Y. Jim\'enez-Teja$^{1,2}$\thanks{E-mail: yojite@iaa.es}, N. Ben\'itez$^2$\thanks{Visiting scientist at the Observat\'orio Nacional}, A. Molino$^{2,3}$ \& C. A. C. Fernandes$^1$ \\
$^1$Observat\'orio Nacional, Rua General Jos\'e Cristino, 77 - Bairro Imperial de S\~ao Crist\'ov\~ao, 20921-400, Rio de Janeiro, Brazil \\
$^2$Instituto de Astrof\'isica de Andaluc\'ia (CSIC), Glorieta de la Astronom\'ia s/n, Granada, E-18008, Spain \\
$^3$Instituto de Astronomia, Geof\'isica e Ci\^encias Atmosf\'ericas, Universidade de S\~ao Paulo, Cidade Universit\'aria, 05508-090, \\
 S\~ao Paulo, Brazil}
\begin{document}

\maketitle

\begin{abstract}
Photometric redshifts, which have become the cornerstone of several of the largest astronomical surveys like PanStarrs, DES, J-PAS or the LSST, require precise measurements of galaxy photometry in different bands using a consistent physical aperture.  This is not trivial, due to the variation in the shape and width of the Point Spread Function (PSF) introduced by wavelength differences, instrument positions and atmospheric conditions. Current methods to correct for this effect rely on a detailed knowledge of the PSF characteristics as a function of the survey coordinates, which can be difficult due to the relative paucity of stars tracking the PSF behaviour. Here we show that it is possible to measure accurate, consistent multicolour photometry {\it without} knowing the shape of PSF. The Chebyshev-Fourier Functions (CHEFs) can fit the observed profile of each object and produce high signal-to-noise integrated flux measurements unaffected by the PSF. These total fluxes, which encompass all the galaxy populations, are much more useful for Galaxy Evolution studies than aperture photometry.  We compare the total magnitudes and colours obtained using our software to traditional photometry with SExtractor, using real data from the COSMOS survey and the Hubble Ultra Deep Field. We also apply the CHEFs technique to the recently published Extreme Deep Field and compare the results to those from ColorPro on the HUDF. We produce a photometric catalogue with 35732 sources (10823 with S/N$\geq$5), reaching a photometric redshift precision of 2\% due to the extraordinary depth and wavelength coverage of the XDF images.
\end{abstract}


\section{Introduction}\label{intro}

 The accuracy in photometric measurements depends on many factors, e.g. the precision in modelling the Point Spread Function (PSF) and its variation across the image, the proper determination of the background and its variance, the definition of the aperture, etc. SExtractor \citep{sextractor,sextractor2}, the SDSS photometric pipeline \citep*{sdss,lupton} and ColorPro \citep{dan} are among the most successful and widely used codes for astronomical photometry, although their approaches are very different. SExtractor removes the background and estimates the flux of the sources using different kinds of apertures, for instance, circular, elliptical or isophotal apertures. Despite its versatility and usefulness, SExtractor suffers from certain problems, since it tends to underestimate the background noise and miss the wings of extended objects, leading to an underestimate of the flux and its error \citep{sesar,molino}. Also, when calculating colours, aperture photometry is measured in different filters degraded to the worst PSF in order to measure a constant fraction of the total light in every band and to avoid artificial colour gradients. This degradation causes the loss of valuable information.\\
 
 The SDSS survey provides several measurements of galaxy photometry for different sciences goals. For instance, the SDSS photometric catalogue provides Petrosian magnitudes (which recover the flux within a certain radius, thus providing a fraction of the total flux), adequate for nearby, bright galaxies. Model magnitudes are computed by fitting the objects with traditional profiles, such as de Vaucoulers or exponential functions convolved with the PSF. This model photometry yields total magnitudes, although \citet{lupton} caution that there may be inaccuracy for complex galaxies with significant substructure, which due to the relatively poor quality of the SDSS imaging, it is not the case for most SDSS objects. Colours are measured within a certain aperture, calculated from a simple profile that is applied to all the different filters. Thus, although very complete and successful for the Sloan data, this photometric pipeline is bound to have problems when dealing with higher S/N data which resolve more details of complex galaxy profiles.\\
 
 ColorPro obtains aperture-matched, PSF-corrected photometry without degrading the quality of the images. It uses a linear combination of the SExtractor MAG\_AUTO and MAG\_ISO magnitudes for each filter and the detection image. The result is a highly robust, precise, and unbiased photometry that has been successfully applied to important large surveys, such as ALHAMBRA \citep{moles, molino} or CLASH \citep{clash}. Again, as in the two previous cases, this technique requires knowledge of the PSF, which in many cases becomes a difficult, time-consuming task. Moreover, the current version of ColorPro is not capable of dealing with variability of the PSF across the image since the photometric corrections are calculated globally, degrading the whole detection image to the seeing of each individual filter. This is not an issue with relatively small fields, but it will hinder its application to the new very large surveys.\\
 
 When using models to fit galaxies before computing the photometry, as in the SDSS, the final result will be highly affected not only by the factors mentioned at the beginning of this section but also by the precision of the fit. Modelling techniques are traditionally classified in two groups: parametric and non-parametric methods. The first approach includes algorithms as the above-mentioned SDSS photometric pipeline, GALFIT \citep{galfit1,galfit2} or GALAPAGOS \citep{galapagos}. The latter two are also based on simple analytical profiles, allowing the simultaneous fit of several of them. Parametric techniques mostly use profiles roughly similar to the shape of the galaxies, with parameters which are associated to physical characteristics, and due to their low complexity, are straightforward to implement and optimize. However, the selection of the profiles and the determination of the initial parameter values requires intervention from the user, making these techniques less suitable for processing the huge amount of data coming from large surveys. To adapt them for this purpose, it is necessary to highly restrict the parameter space, to avoid them being ill-posed. This even further limits their flexibility to model the wide variety of morphologies observed in the sky and thus does not accurately represent irregular objects or the substructures present in complex and well resolved galaxies. For these reasons, the photometry provided by parametric fits is often inaccurate. \\
 
 The non-parametric techniques \citep{shapelets,sersiclets,sinh} generally use orthonormal bases to model objects, which a priori makes them capable of fitting any kind of morphology. These methods are purely mathematical, so the user does not need any a priori information and the algorithms are fully automated. They have the drawback of having coefficients whose physical meaning is not easy to interpret. Moreover, these techniques are usually based on continuous and infinite functions, and the transition to the discrete and finite images is not trivial. In spite of these disadvantages, models obtained by these methods tend to be more precise than parametric ones, and thus yield better photometric measurements.\\
 
 The CHEFs (Chebyshev-Fourier bases) \citep{yoli} constitute a non-parametric fitting technique specially developed to efficiently model any kind of galaxy morphology with a very compact and accurate decomposition. As we show here, they have been applied to the measurement of total magnitudes overcoming most of the problems described above for the previous methods and thus yielding highly precise multicolour photometry without requiring PSF measurements, saving a considerable amount of time and effort, as well as leading to a completely automated algorithm, easily applicable to large surveys pipelines. The CHEFs will be one of the main tools to yield the precise photometry necessary for the upcoming J-PAS survey \citep{jpas}.\\
 
 The higher the resolution and S/N of the images, the more difficult it is for both parametric and non-parametric methods to achieve accurate fits. That is why we use the HUDF to test the reliability of the CHEF photometry. By using real data from the HUDF and comparing the results with those of SExtractor, we prove that our photometry is unbiased and extremely accurate. We have also applied our technique to the new, very deep data from the Extreme Deep Field \citep[XDF,][]{XDF} to compare the results with those from \cite{dan} and to obtain the first public catalogue for this image.\\
 
 This paper is organized as follows: we briefly describe the CHEF mathematical background in Sect. \ref{chef_phot} as well as the photometric pipeline. Sect. \ref{test} is devoted to testing the CHEF photometric measurements and comparing them to the ones provided by SExtractor, using a set of extremely realistic models from the HUDF as a sample test. In Sect. \ref{sect_color} we extend our test to the colours, comparing CHEFs with traditional aperture photometry. In Sect. \ref{XDF_section} we apply the CHEFs to the recently released data from the XDF, yielding the first photometric catalogue for it. We also compare the accuracy of the photometric redshifts obtained by CHEFs against those from ColorPro.\\

\section{CHEF photometric pipeline} \label{chef_phot}

 In \citet{yoli}, we presented an analytical formula to measure the flux just using the CHEF coefficients. Now we implement this calculation into an algorithm which maximizes the capabilities of the CHEF decomposition and incorporates the effect of the PSF without computing it nor any kind of correction. The basic idea is to evaluate the CHEF models within a frame large enough to enclose all the flux from a galaxy, determining the radius up to which the object extends and measuring its total magnitude. This will provide a PSF-independent measurement of the magnitude, which can be compared across filters, avoiding the time-consuming determination of the PSF and the possible errors introduced by it.\\
 
\subsection{Mathematical background}

 Every bidimensional, smooth, square-integrable function $f$  can be fit using the CHEF bases \citep{yoli}, which are composed by Chebyshev rational functions and trigonometric waves in polar coordinates:
\begin{equation}\displaystyle
\left\{\phi_{nm}(r,\theta;L)\right\}_{nm} =\left\{\frac{C}{\pi}\;TL_{n}(r,L) 
W_{m}(\theta)\right\},\label{basis}
\end{equation}

\noindent with $TL_n$ the Chebyshev rational function of order $n$ \citep{boyd}, which depends on the radial coordinate $r$ and the scale factor $L$ related to the speed of the functions in reaching the extrema. $W_{m}(\theta)$ is a general
expression to represent both $\sin{(m\theta)}$ and $\cos{(m\theta)}$, and $C$ a normalization factor to make the final functions orthonormal, with value $\displaystyle C=\left\{
\begin{array}{ll} 1, & \mbox{if } n=0 \\ 2, & \mbox{if } n>0 \end{array}\right.$. Please notice this basis is indexed by two indices $n$ and $m$ which refer to the order of the Chebyshev polynomial that each basis function comes from and to the Fourier frequency considered, respectively. So $f$ can be decomposed into the linear combination of these basis functions and some coefficients $f_{nm}$, the so called {\it CHEF coefficients}. These coefficients are calculated by the weighted inner product
\begin{equation}
\displaystyle f_{nm}=\frac{C}{2\pi^2}\int\limits_{-\pi}^\pi
\int\limits_0^{+\infty} f\left(z,\phi\right)TL_{n}\left(z\right)W_m\left(\phi\right)
\frac{1}{z+L}\sqrt{\frac{L}{z}}\;dz\;d\phi.\label{coeffs}
\end{equation}

\noindent The weight is needed to ensure the orthogonality of the basis. It must be noticed that every function $f$ will be best fitted using an optimal value for the scale parameter $L$. Compact galaxies need smaller values of $L$ whereas extended galaxies are better represented by larger scale parameters. To perform photometric measurements, we also derived in \citet{yoli} a formula to calculate the flux enclosed by a circular aperture of radius $R$, just by using the CHEF coefficients associated to the cosine functions with Fourier frequency $m=0$, since all the other terms are cancelled during the integration process:
\begin{equation}F(R)=\int\limits_0^{R}\int\limits_{-\pi}^\pi f(r,\theta)r\; d\theta dr=2\pi \sum_{n=0}^{+\infty} f_{n,0}^c\,I_1^{n}.\label{flujo}\end{equation}

\noindent In this expression the superindex $c$ indicates the CHEF coefficient related to the cosine and $I_1^n$ is the result of the integration of the Chebyshev rational functions in a circular area of radius $R$:
\begin{equation}\begin{array}{rcl}
I_1^{n}&=&\displaystyle 2\sum_{j=0}^{n} \left(\begin{array}{c} n \\ j \end{array}\right) (-1)^j L^{-j/2}\frac{R^{j/2+2}}{j+4}\cdot \\
& & \displaystyle \cdot Re\left(e^{in\pi/2}i^{n+j} \,_2F_1\left(n,j+4,j+5;\frac{-i\sqrt{R}}{\sqrt{L}}\right)\right),\end{array}\label{integral1}
\end{equation}

\noindent where  $\,_2F_1$ stands for the hypergeometric function in the case $n>0$. If $n=0$, then simply
\begin{equation}I_1^{0}=\frac{R^{2}}{2}.\label{integral2}\end{equation}

 In short, taking advantage of the CHEF compactness, it is possible to calculate the flux inside a circular aperture using just a few coefficients, all of them with zero angular component.\\

\subsection{Practical implementation} \label{practical}

 As described in \citet{yoli}, the CHEF pipeline is implemented as a Python code which automatically selects the optimal number of coefficients $n$ and $m$ to be computed as well as the optimal value for the scale parameter $L$ (see expression (\ref{basis})), which was originally closely related to the half-light radius of the galaxy. During the implementation of the CHEF photometric pipeline for this current work, we realized that the SExtractor estimation of this half-light radius is unstable for faint galaxies. Hence we preferred to substitute the value of the scale parameter $L$ by another quantity that unambiguously represents the size of an object. We defined $L$ as the radius of a circular aperture such that its area is equivalent to the isophotal area provided by SExtractor. This parameter $L$ is used not only to define the slope of the CHEF basis functions but also to determine the extension of the postage stamp in which each object is modelled. The aim is to make it large enough to catch all the light from the source but also small enough to avoid excessive background noise contamination. We find that a good compromise is setting the stamp radius equal to $2L$. That means that each CHEF model has an area four times larger than the SExtractor isophotal area, which should be more than enough for our purposes. This modification turned out to be a great improvement in the accuracy achieved by the CHEF models (especially for faint magnitudes), with more accurate fits using a lower number of coefficients. It should also be stressed that to get good photometry is not essential to have the exact optimal value for the scale parameter $L$, since the CHEF bases are flexible enough to get very accurate models compensating the imprecision in $L$ with the number of Chebyshev and Fourier coefficients $n$ and $m$.\\
 
 Once we set the stamp size and evaluate the CHEF model inside it, it is necessary to find an objective criteria to define the real extension of the galaxies (that is the limit $R$ of the integral (\ref{flujo})), excluding those areas where the noise dominates the signal. For each object, the algorithm calculates the radial flux of the CHEF model (free of noise) and determines the radius where it converges (that is, where the flux reaches a maximum or oscillates by less than 1\% of the total flux). This is considered to be the aperture radius containing the full flux of the object, so no corrections for the PSF have to be applied to compare across different filters with different PSFs. Hence, the correction for the PSF is not needed to get the colours. Using this radius $R$ we integrate the CHEF model using the eqs. (\ref{flujo}), (\ref{integral1}), and (\ref{integral2}), to obtain its total flux. Note that different apertures $R$ can be used for each image an object is observed in.\\ 
 
\begin{figure}\centering
\includegraphics[width=8.5cm]{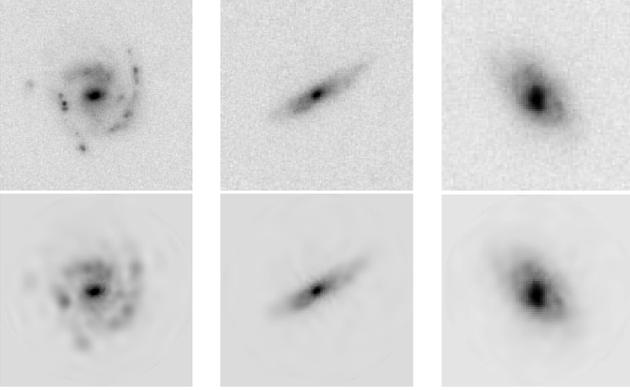}
\caption{Examples of galaxies modelled with CHEFs. We selected three real galaxies from the XDF with different morphologies to show the efficiency of CHEFs at recovering the radial profiles and thus defining the total extension of the objects. The top row shows the original galaxies and the bottom the resulting CHEF models. These three galaxies correspond to the ones used in Figure \ref{profiles}. The images have been normalized and are shown at the same scale. We refer the reader to \citet{yoli} for more examples.}\label{exgalaxies}
\end{figure}

\begin{figure}\centering
\includegraphics[width=8.5cm]{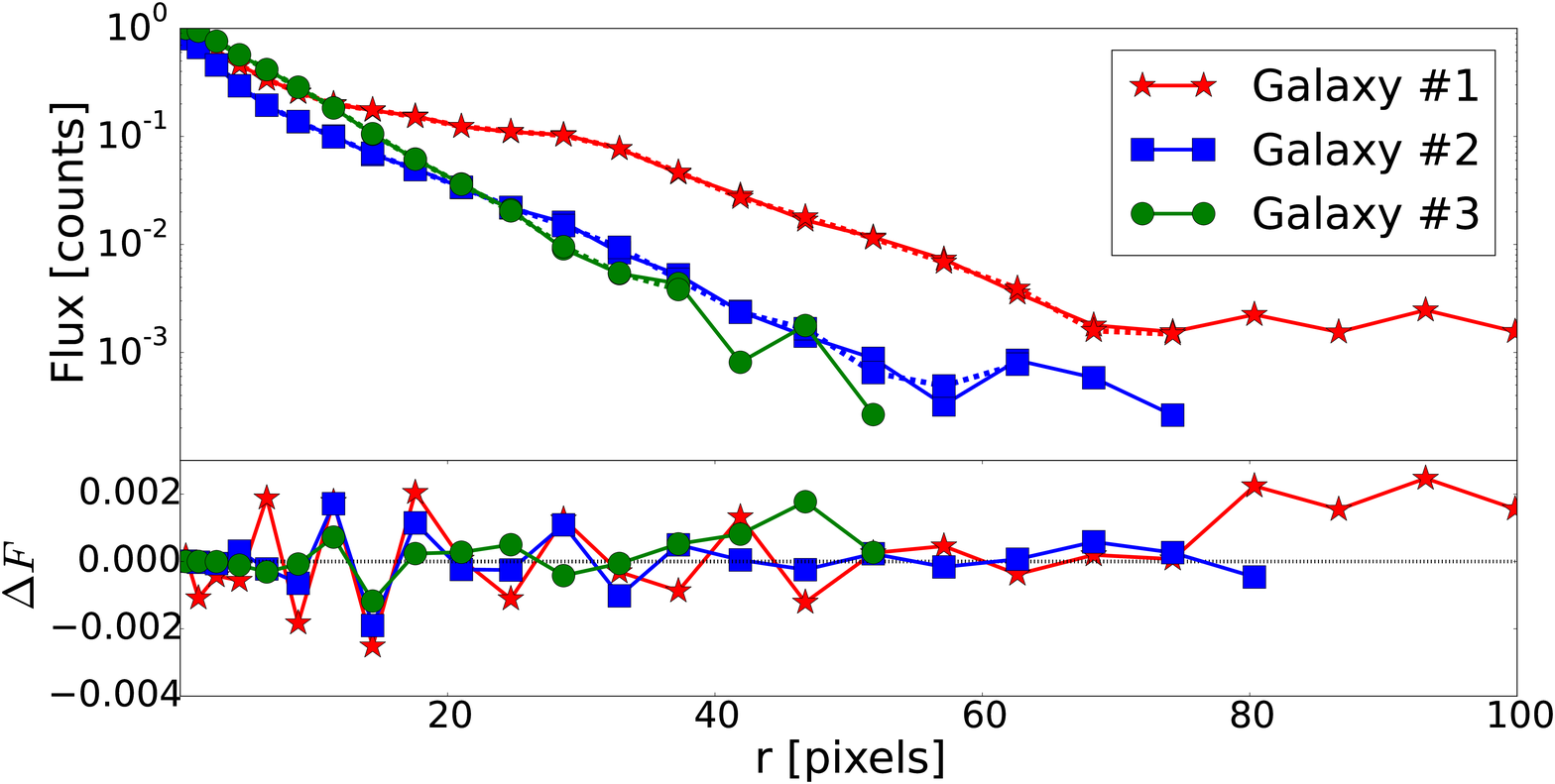}
\includegraphics[width=8.5cm]{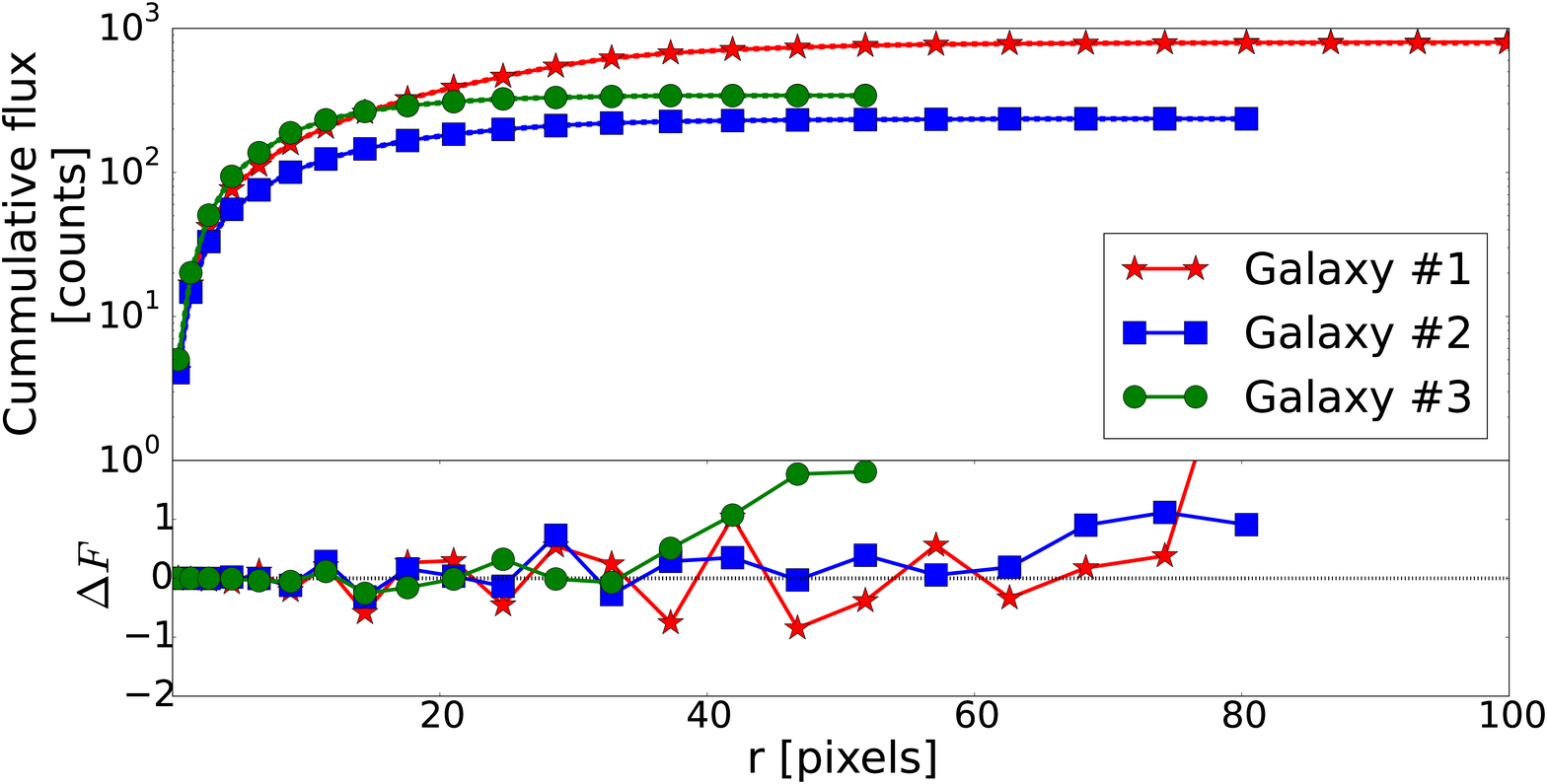}
\caption{Radial and cumulative radial profiles for the original galaxies and their CHEF models, shown in Figure \ref{exgalaxies}. The profiles are plotted in the top panel of each figure, on a logarithmic scale. Solid lines represent the original profiles and the dotted lines are the CHEF profiles. The bottom panels display the linear flux difference between the original data and the CHEF models. As the images are normalized, the difference in the radial profiles (top panel) expresses the percentage of error achieved. [\textit{See the electronic edition of the
journal for a colour version of this figure.}]}\label{profiles}
\end{figure}

 To see how the CHEF flux with increasing radius is compared to traditional aperture photometry in increasing apertures we have randomly chosen three galaxies from the XDF \citep{XDF} with different morphologies and calculated both the radial and the cumulative radial profiles. Figure \ref{exgalaxies} shows the three galaxies we chose to illustrate the CHEFs performance, along with their corresponding CHEF models. Fig. \ref{profiles} shows that the resulting curves for the CHEF models (dotted lines) and the original galaxies (solid lines) are almost undistinguishable. This conveys the efficiency of the CHEFs not only fitting different morphologies, but also determining the radius $R$ of the galaxies using the above described method.\\
 
 The error in this magnitude is calculated by means of the expression \citep{newberry}
 
 \begin{equation}
 err_m=1.0857\sqrt{\pi\left(\frac{R\sigma}{F}\right)^2+\frac{1}{FT_{exp}g}},\label{error}
 \end{equation}

\noindent where $F$ denotes the flux from Eq. (\ref{flujo}), $\sigma$ the standard deviation of the noise, $T_{exp}$ the exposure time of the image and $g$ its gain.\\
 
 Further information on the code function can be found in Section \ref{chefs_settings}.\\

\section{Photometry test} \label{test}

In \citet{yoli} the performance of CHEFs for measuring photometry and shapes was compared to the shapelets technique \citep{shapelets}. A sample of analytical simple profiles (S\'ersic functions with indices ranging from 0.5 to 4 and sheared with different levels of ellipticity), convolved with different PSFs, and blurred with Gaussian noise was used. Under these conditions, CHEFs behaved better than shapelets for measuring both fluxes and shapes, with an homogeneous error independent of the profile and ellipticity of the galaxies and approximately an order of magnitude lower than the shapelets'.\\

\begin{figure}
\centerline{\includegraphics[width=8.5cm]{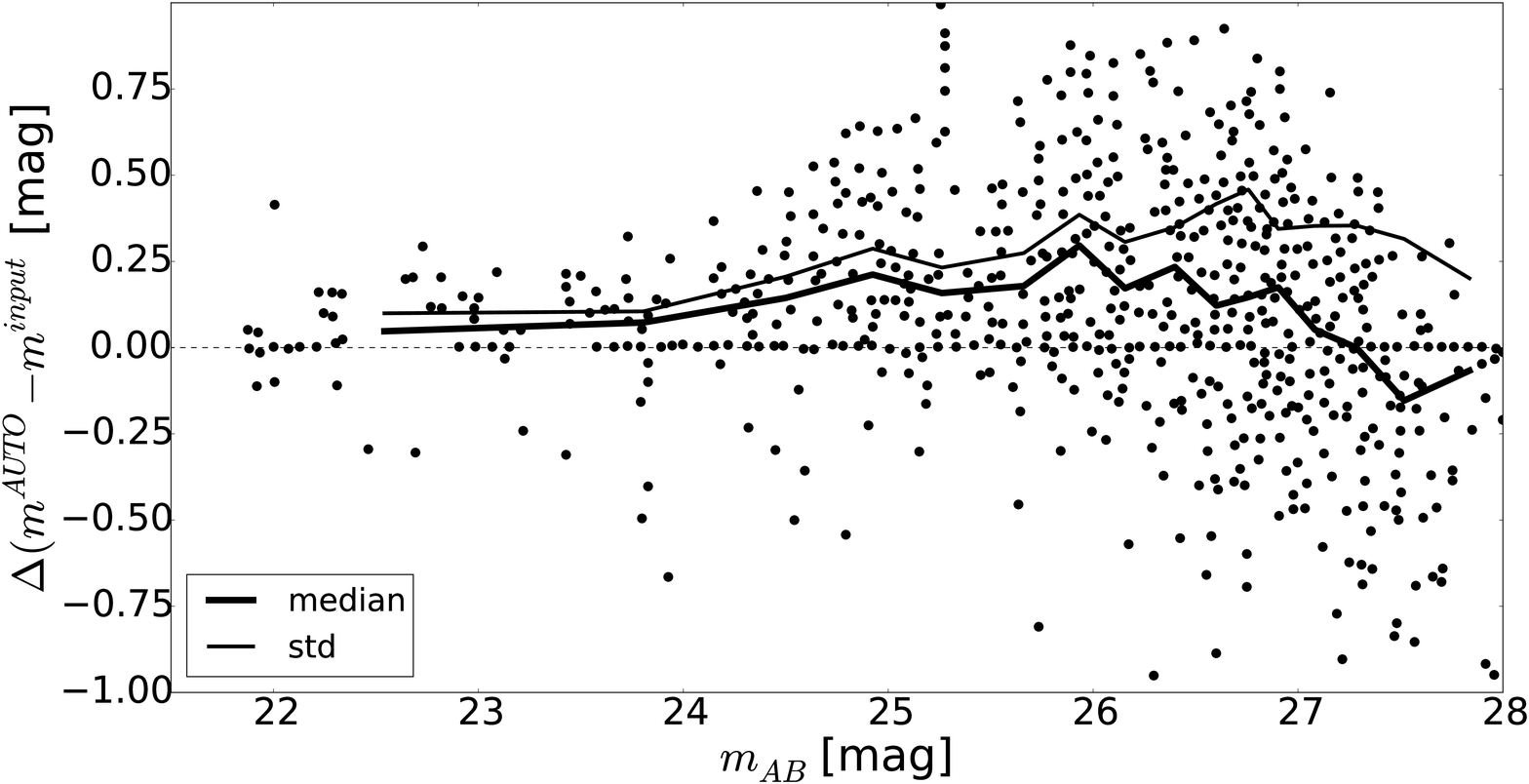}}
\centerline{\includegraphics[width=8.5cm]{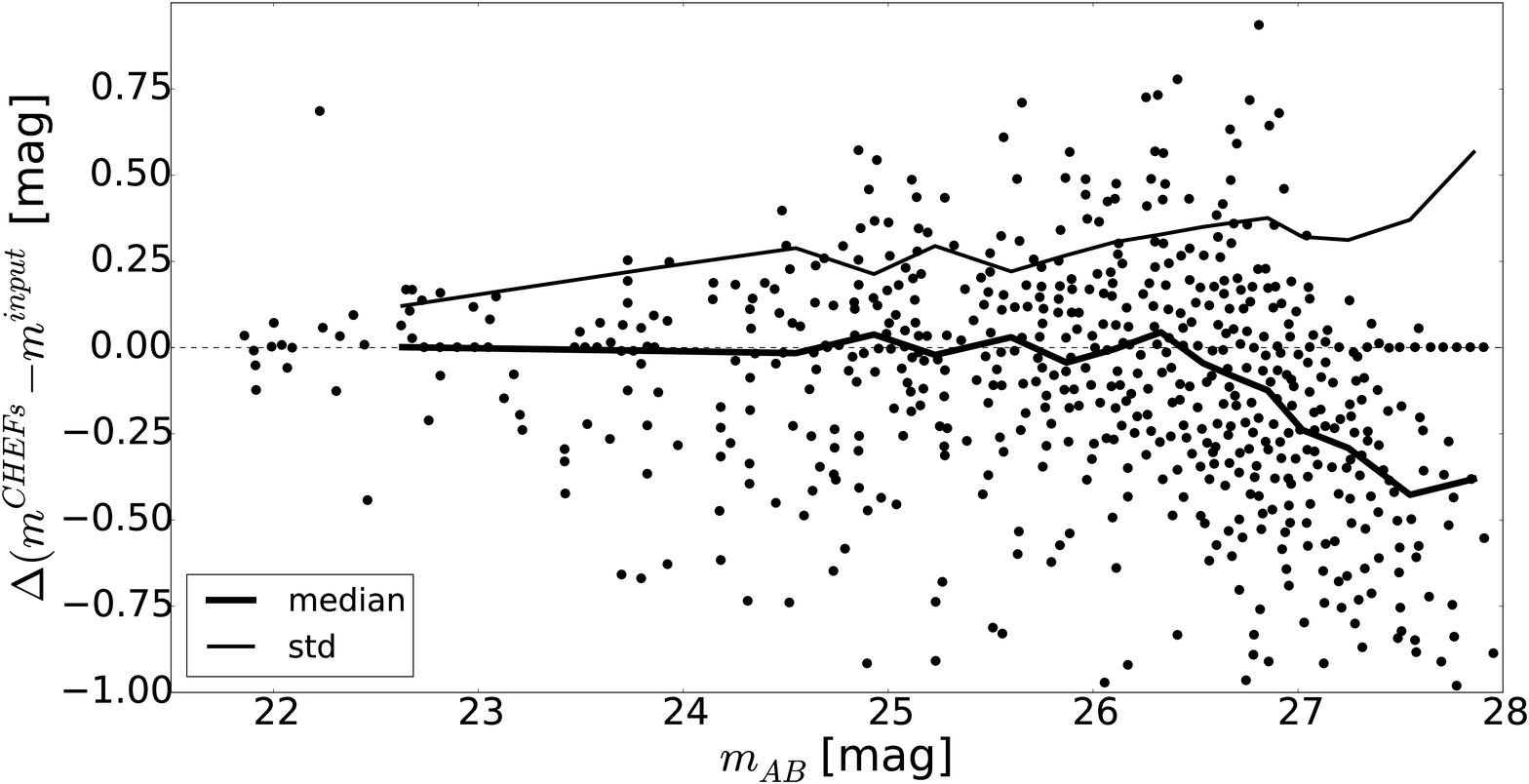}}
\caption{Magnitude comparison for the UDF models inserted in a COSMOS field observed by SUBARU: top panel shows  the difference between the recovered SExtractor MAG\_AUTO and the real analytical magnitude; bottom panel displays the difference between the recovered total magnitude measured by the CHEFs and the real analytical one.} \label{test2}
\end{figure}

In this paper, we have used a more realistic and complex sample of models to test the CHEF photometry against SExtractor. We have taken the real CHEF models from the UDF adapted to the observational characteristics of a ground-based survey, as if they had been observed by SUBARU, i.e., repixelated and convolved with the corresponding PSF but without adding synthetic noise. We have inserted them at random locations in a COSMOS field \citep{cosmos} observed by SUBARU, so that they have the photometric noise of the COSMOS observations. To ensure that we are not biasing our results using these CHEF models as a testing sample, we also rotated the models arbitrarily and recentred them with sub pixel shifting. After rotation, interpolation, PSF convolution, and noise addition, we can be certain that the final coefficients of the objects have no relation with the original ones. We have measured the magnitudes of these degraded objects using both SExtractor MAG\_AUTO parameter, which is the closest to a total magnitude, and the CHEF algorithm described in Sect. \ref{chef_phot}. We have then compared both measurements with the original analytical magnitude, directly obtained from the UDF data. The results can be seen in Figure \ref{test2}, where the comparison of the real magnitudes with the SExtractor and CHEF ones is shown, respectively. We notice that SExtractor tends to underestimate the photometry, displaying a clear bias of $\sim$0.1 magnitudes approximately up to magnitude 24 and larger for fainter sources. This is explained by the small area within the aperture considered by MAG\_AUTO. Since these imprecisions are related to the SExtractor algorithm itself, they do not depend on the parameter configuration selected. However, the CHEFs do not show any kind of bias up to magnitude 26 and the results are much more homogeneous and precise. The scatter in this case is slightly higher, which is completely justifiable since the CHEFs calculate total magnitudes and thus have a larger effective area with more background noise. Notice this scatter gathers the contribution of both the imprecision of the CHEF model and the area-dependent photometric noise.\\

\section{Colour test} \label{sect_color}

 The improvement of CHEFs over SExtractor for recovering total magnitudes is demonstrated in Section 3. Here we test the accuracy of both techniques in measuring colours. With this purpose we used again the i-band image from the UDF and degraded it to the observational characteristics of a ground-based survey, in this case, ALHAMBRA \citep{moles,molino}. We later convolved this degraded image with four different PSFs, Moffat profiles with index 3 and FWHMs ranging from 0.6 to 1.2, which are close to the real ALHAMBRA PSFs \citep{molino}. With this procedure we aimed to obtain something similar to a degraded UDF image in the $bviz$ bands, but without interpolating the original data in wavelength and morphology. Thus, the final colours in these bands should be zero and the test would easily show us the efficiency of the photometric techniques. Given that the CHEFs calculate total magnitudes, we can directly obtain the colours without applying any PSF correction. However, as SExtractor does not offer the same possibility, we applied the widely-used technique of degrading our images to match the one with the worst PSF and measuring the photometry in 3 arcsec apertures afterwards, running SExtrator in dual-image mode. We have chosen this aperture size because once all the bands are reduced to the same PSF frame, all the images have a common PSF of 1.2 arcsec. Taking an aperture of this size would be optimal in terms of S/N to measure the photometry of the stars. However, as we are dealing with extended sources, this aperture would be too small and we would be underestimating the photometry, as in the previous section. The distribution of the FWHMs of the objects in the {\it z} band, is a Gaussian centred on 2.8 arcsec. Considering that the pixel scale of the ALHAMBRA images is 0.221"/pix, that means one pixel smaller than the 3 arcsec aperture. Moreover, 68\% of the galaxies in this image (the $2\sigma$ of the distribution) have a FWHM equal or smaller than 3 arcsec. Hence this aperture represents a good trade-off between the most extended sources and those which would not require a such a large aperture.\\
 
\begin{figure}\centering
\centerline{\includegraphics[width=9.5cm]{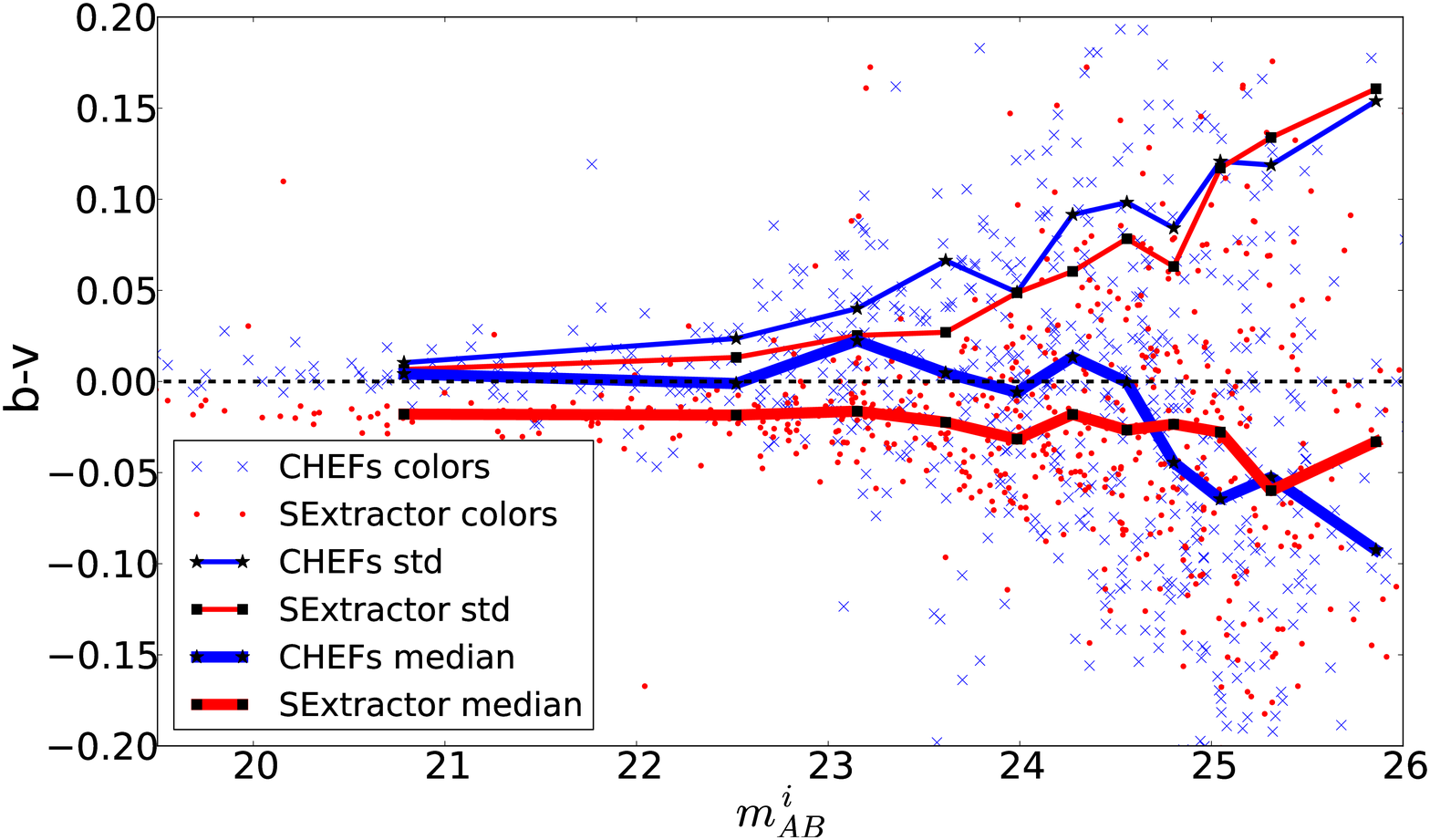}}
\centerline{\includegraphics[width=9.5cm]{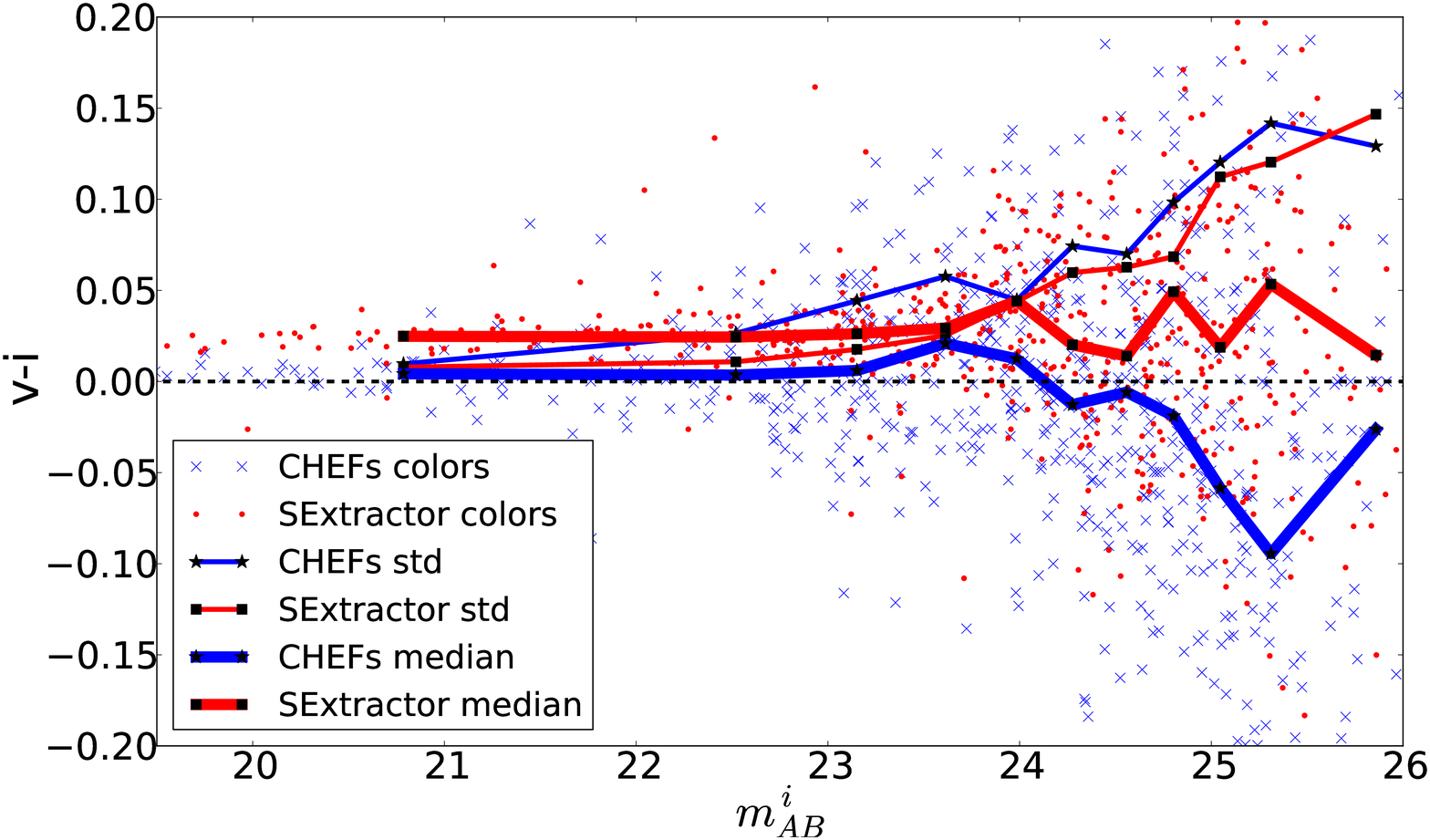}}
\centerline{\includegraphics[width=9.5cm]{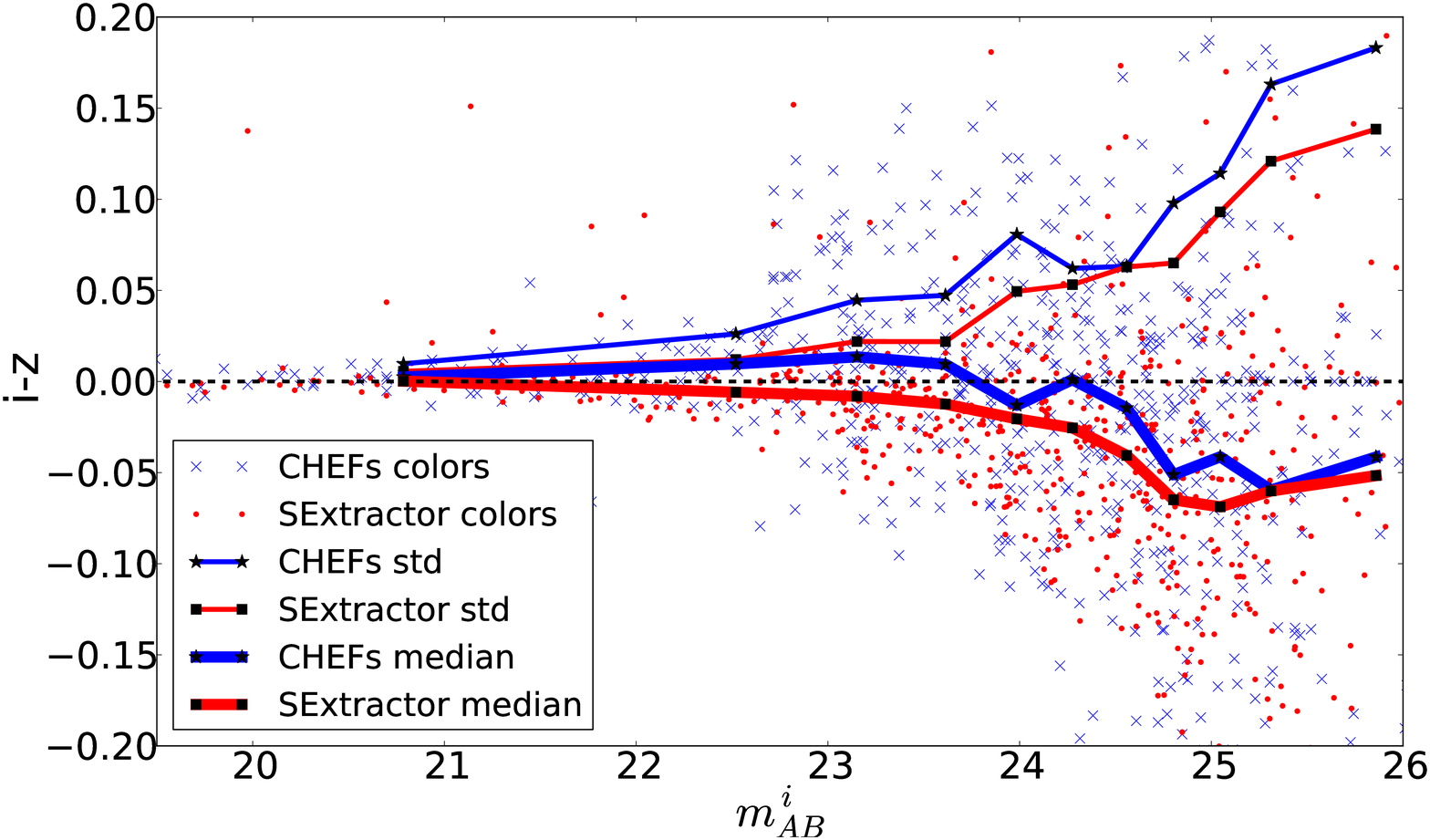}}
\caption{CHEF versus SExtractor colours. Four synthetic images were created from the same UDF i-band, using four different PSFs. The colours obtained by both the CHEFs (blue crosses) and SExtractor (red points) for the three independent colours b-v, v-i, and i-z are displayed from top to bottom. It is also shown the median (thick lines) and the standard deviation (thin lines) of the results by the CHEFs (starred blue lines) and SExtractor (squared red lines). [\textit{See the electronic edition of the journal for a colour version of this figure.}]} \label{testcolor}
\end{figure}

 Figure \ref{testcolor} summarizes the results. The three independent colours $b-v$, $v-i$, and $i-z$ calculated for all the objects detected in these four bands are displayed. It is clear that SExtractor measurements remain biased, as previously observed for AUTO magnitudes. The bias is noticeable for $b-v$ and $v-i$ colours up to magnitude 25 and 24, respectively. The $i-z$ colour does not exhibit this behaviour, since the $z$-band is the only one not degraded in the process of matching the different PSFs, and thus less distorted during the measurements. For this particular colour, the efficiency of the CHEFs and SExtractor is similar, with a slightly better performance of the former for faint objects. However, in $b-v$ and $v-i$ colours the improvement of our method over SExtractor is clearer, since the CHEFs are unbiased up to high magnitudes. The scatter is comparable for both techniques although slightly lower for SExtractor, given that the CHEFs calculate total magnitudes and the measurement is done over a larger area, as described above.

\section{XDF catalogue: redshift test} \label{XDF_section}

 The Extreme Deep Field \citep{XDF} is the deepest area in the Universe ever imaged, as a result of the combination of the optical and near-infrared data of the Ultra Deep Field collected by the Hubble Space Telescope during ten years of work and nineteen observation programs. It provides a superb opportunity for studying very faint objects, with a typical depth of 30 AB mag in most filters. Covering 2 arcmin$^2$ less than the HUDF, that is, 10.8 arcmin$^2$, in 9 filters (5 optical with the ACS and 4 near-IR with the WFC3), these nine bands add up to a total exposure time of 21.7 days. Both the photometric depth and the high S/N of this image make this field an unique dataset to show the performance of the CHEFs. The analysis consists of three basics steps, that can be summarized in: 1) obtaining suitable SExtractor configuration parameters to detect the sources in the image, 2) decomposing the detected objects into CHEF components, and 3) running BPZ \citep{bpz} to calculate the photometric redshifts. We finally compare our results to the ones published in \cite{dan}.\\
 
\subsection{SExtractor settings} \label{sect_sex}

 Following the same strategy described by \cite{XDF}, we created a detection image by coadding all the optical filters, that is, $B_{435}+V_{606}+i_{775}+I_{814}+z_{850}$. The unusual depth and exposure time of these data make the field contain very inhomogeneous objects, some of them highly extended and thus covering many other smaller, foreground and background galaxies. As the CHEFs are not able to carry out any detection in the image but just model the objects present in it, we require a detection software. The CHEF code uses SExtractor software \citep{sextractor,sextractor2} not only to perform this task but also to calculate some parameters that the CHEF algorithm needs. The precision, speed and simplicity of SExtractor yield it the most suitable to be integrated in the CHEF pipeline. However, finding a perfect SExtractor configuration file that detects all the objects in the XDF coadded image is not straightforward, since extended galaxies can result deblended if the parameters favour the search for small objects. Nevertheless,  the compact, faint sources can remain undetected if the configuration is settled to efficiently find the largest, brightest galaxies. For that reason, we explored the complete parameter space of SExtractor configuration to find out the set of values which maximized the number of detections with $S/N>5$ (calculated using Kron apertures, with Kron parameters PHOT\_AUTOPARAMS = 1.6, 2.5, as chosen by \cite{XDF}) and minimized the number of spurious detections \citep{molino}.\\ 
 
 As a result of this analysis, we finally required each detection to have 4 contiguous, 1.0$\sigma$ pixels above the background, globally estimated with BACK\_SIZE = 32 and BACK\_FILTERSIZE = 3. The resulting deblending parameters were DEBLEND\_NTHRESH = 32 and DEBLEND\_MINCONT = 0.005. We visually inspected the segmentation-map, ensuring the extended galaxies were properly detected and not split.\\
 
\begin{figure}\begin{center}
\includegraphics[width=8.5cm]{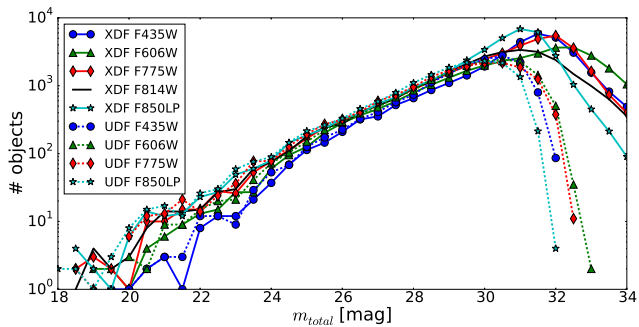}
\caption{Cumulative distribution of the number of galaxies as a function of the total magnitude for the XDF (solid lines) and the UDF (dashed lines), on a logarithmic scale. Total magnitudes are calculated by the CHEFs for the XDF and by ColorPro in the case of the UDF. Colour and marker code identifies the bands. [\textit{See the electronic edition of the journal for a colour version of this figure.}]} \label{num_cuentas}
\end{center}\end{figure}

 With this configuration, we detect a total number of 35732 objects, 10823 of which have $S/N \geq 5$ according to the values of the Kron parameters mentioned above. The catalogue by \cite{dan} for the UDF contains a total of 18633 sources above this threshold (notice the area covered by the UDF is larger, and the definition of S/N is different in this case, since it is calculated using the photometry and photometric errors provided by ColorPro). Figure \ref{num_cuentas} shows the number of sources detected for both the XDF and the UDF on each band as a function of the corresponding total magnitude. Magnitudes were calculated either by the CHEFs or by ColorPro, according to the case. Please notice that just the nearest common bands for the UDF and the XDF are displayed for the sake of a fair comparison, using the same colours and markers for the same filters. Solid lines represent the detections in the XDF, while the dashed ones are referred to the UDF. The number counts peak at magnitude 30.5 for F775W filter in the UDF, whereas the peak falls at magnitude 32 for the XDF. Considering just the sources with $S/N>5$ not only the number of detections but also the peak of the distributions nearly coincide.\\
 
\subsection{CHEFs settings} \label{chefs_settings}

 Once we had a single catalogue and segmentation-map comprising the information about all the objects in the detection image, we run the CHEF code in ``dual-image mode'' (ie. CHEFs call SExtractor in dual-image mode), and used the output information to analyze the data of the nine optical and near-IR filters. The CHEFs firstly sort the objects by MAG\_AUTO magnitude and decompose them one by one, subtracting every CHEF model before fitting the following one. The CHEF algorithm borrows some other parameters from SExtractor, as for instance, the centroid or the number of pixels in the ISOAREA\_IMAGE (to define the scale parameter $L$ for the CHEF basis functions, as explained in Sect. \ref{practical}). Notice this parameter is independently evaluated for each band and does not come from the detection image. Therefore, the same object will have a different scale $L$ depending on the filter and its models will be computed within circular stamps of different size given that the radius is set to $2L$. It is also important to remark that the same object can exhibit different shapes when observed by different filters, thus introducing artificial colour gradients in aperture photometry.\\
 
 To decompose each object, the CHEFs firstly subtract the background corresponding to that stamp. Two possibilities are available for this step, letting the user choose between using the background-map provided by SExtractor or locally estimating the background and its noise using an internal algorithm. This algorithm, as SExtractor does for non-crowded fields, computes the background as the mean of the pixels on the stamp that do not belong to any source (information provided by the SExtractor segmentation-map), after iteratively discarding all those pixels above the $3\sigma$ level. The difference with SExtractor is that we do not just estimate the background noise scaling the pixel-to-pixel noise with a formal Gaussian, thus assuming a Poissonian background noise distribution at large scales, but find the empirical relation of the noise with the aperture size. The standard image processing (dithering, degradation, stacking,...) introduces correlations between neighbouring pixels making the background noise in images different from a Poissonian distribution. Thus, for many images, the SExtractor background noise estimation is not accurate enough and it is preferable to use the CHEF internal algorithm.\\
 
\begin{figure}
\centerline{\includegraphics[width=7cm]{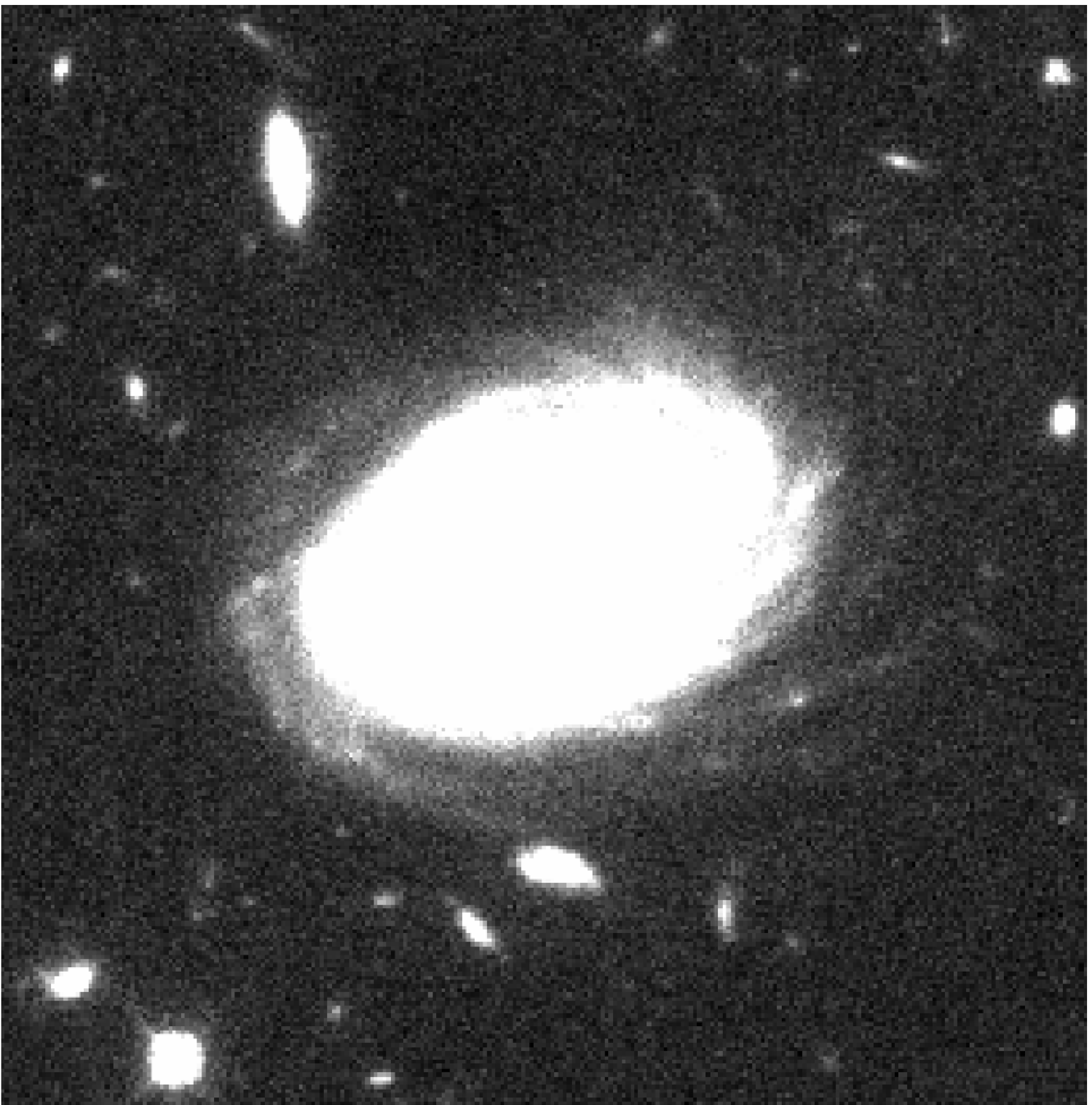}}
\centerline{\includegraphics[width=7cm]{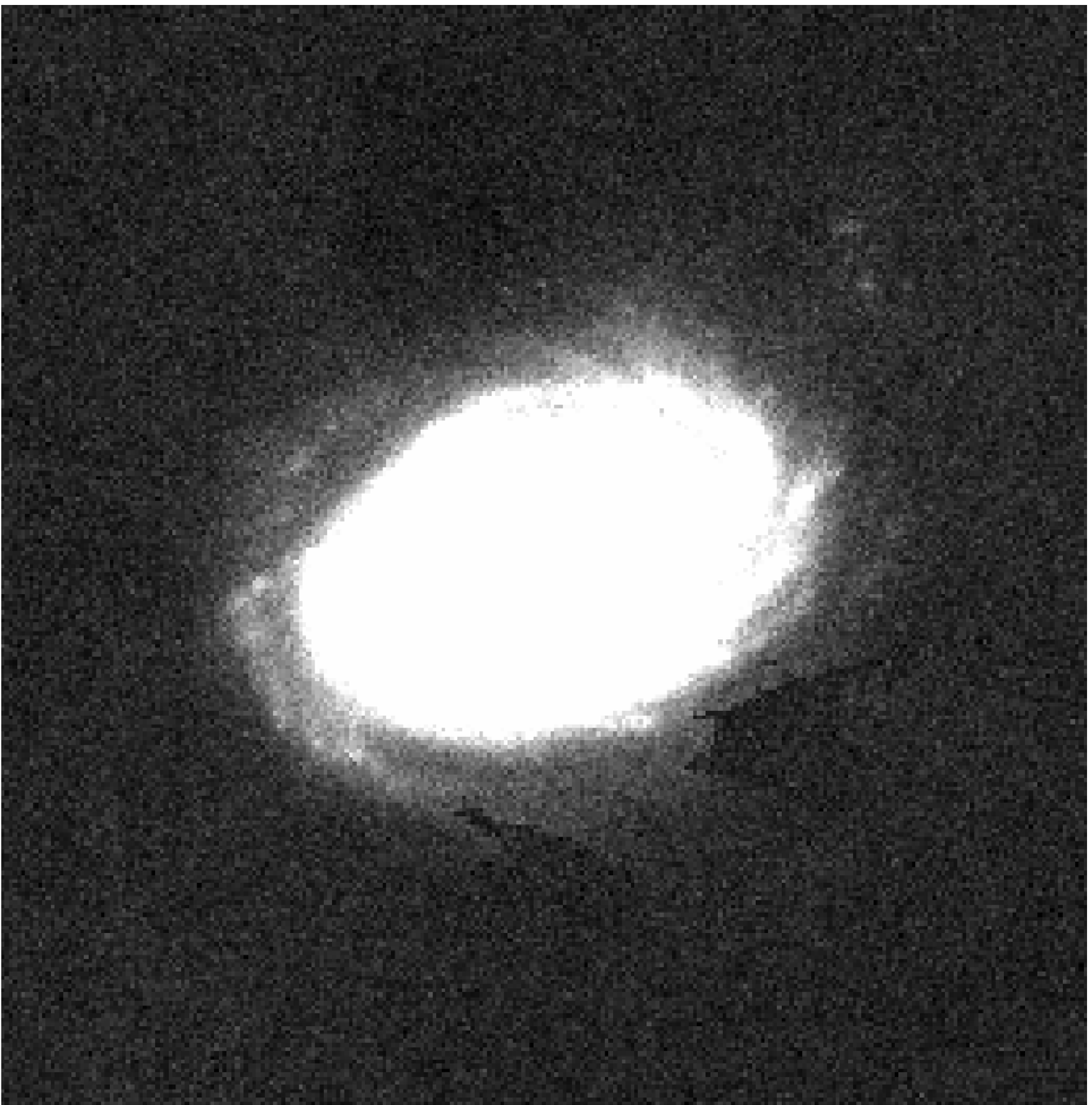}}
\caption{Extended galaxy from the XDF, before and after masking the neighboring objects using the SExtractor segmentation map calculated as explained in Section \ref{chefs_settings}. These objects are replaced by an average of the surrounding pixel and Gaussian noise according to the background RMS.} \label{galaxy_masked}
\end{figure}

 After this background subtraction, the neighbouring, fainter sources present in the stamp are masked, according to the segmentation-map. These objects are substituted by an average of the surrounding pixels plus Gaussian noise equivalent to the background noise in that area. This leads to a smooth, contamination-free main galaxy, whose centroid is the origin of the CHEF functions grid (see Fig. \ref{galaxy_masked} for an example of an extended galaxy from XDF that has been isolated from the closest sources before the CHEF processing). Please note this masking step is only applied to the faintest sources in the stamp, since the brightest ones were previously subtracted by the CHEF algorithm itself in previous iterations.\\
 
 The CHEF code internally calculates the optimal number of CHEF coefficients to model each object, that is, the optimal number of Chebyshev rational functions, $n$, and Fourier frequencies, $m$. However, it is necessary to set the maximum number of coefficients allowed, for the sake of speed. Note that fast algorithms will be necessary to manage the huge amount of data that will be provided by the upcoming large surveys. For the particular analysis of the XDF data, the maximum number of coefficients allowed to compute was set to $n=m=15$ for those objects with postage stamps larger than 100x100 pixels and $n=m=10$ for the remaining. Both the background and the RMS background were internally computed by the CHEF code using its own local estimation. With these settings, the CHEFs took $\sim$ 48 hours to process the 9 bands of the XDF, using an eight-core computer. The values for the photometric zero points and the exposure time were directly extracted from \cite{XDF}.\\
 
 As we were using 1) our own local background noise estimation, which yields larger values than SExtractor as described above, and 2) larger areas for the flux calculations, our magnitude errors calculated by Eq. (\ref{error}) are expected to be higher than those by SExtractor. However, our magnitude errors are comparable to the ones computed by ColorPro \citep{dan} since these include the error of the PSF correction. With all this photometric information we also calculate the $5\sigma$ limiting AB magnitude for each band, summarized on Table \ref{limmag}.\\
 
\begin{table}\centering\begin{minipage}{90mm}\caption{$5\sigma$ limiting AB magnitudes of the nine XDF bands, calculated with the CHEFs photometry.}\label{limmag}
\begin{center}\begin{tabular}{lcc}\hline
Camera & Filter & Depth (AB)\\
 & & [mag]  \\ \hline
WFC3 & F105W &  29.4\\
WFC3 & F125W &  29.1\\
WFC3 & F140W &  27.0\\
WFC3 & F160W &  28.7\\
ACS & F435W &  29.0\\
ACS & F606W &  29.7\\
ACS & F775W &  29.5\\
ACS & F814W &  28.3\\
ACS & F850LP & 28.7\\
\hline
\end{tabular}\end{center}
\end{minipage}\end{table}
 
\begin{figure}
\centerline{\includegraphics[width=8.5cm]{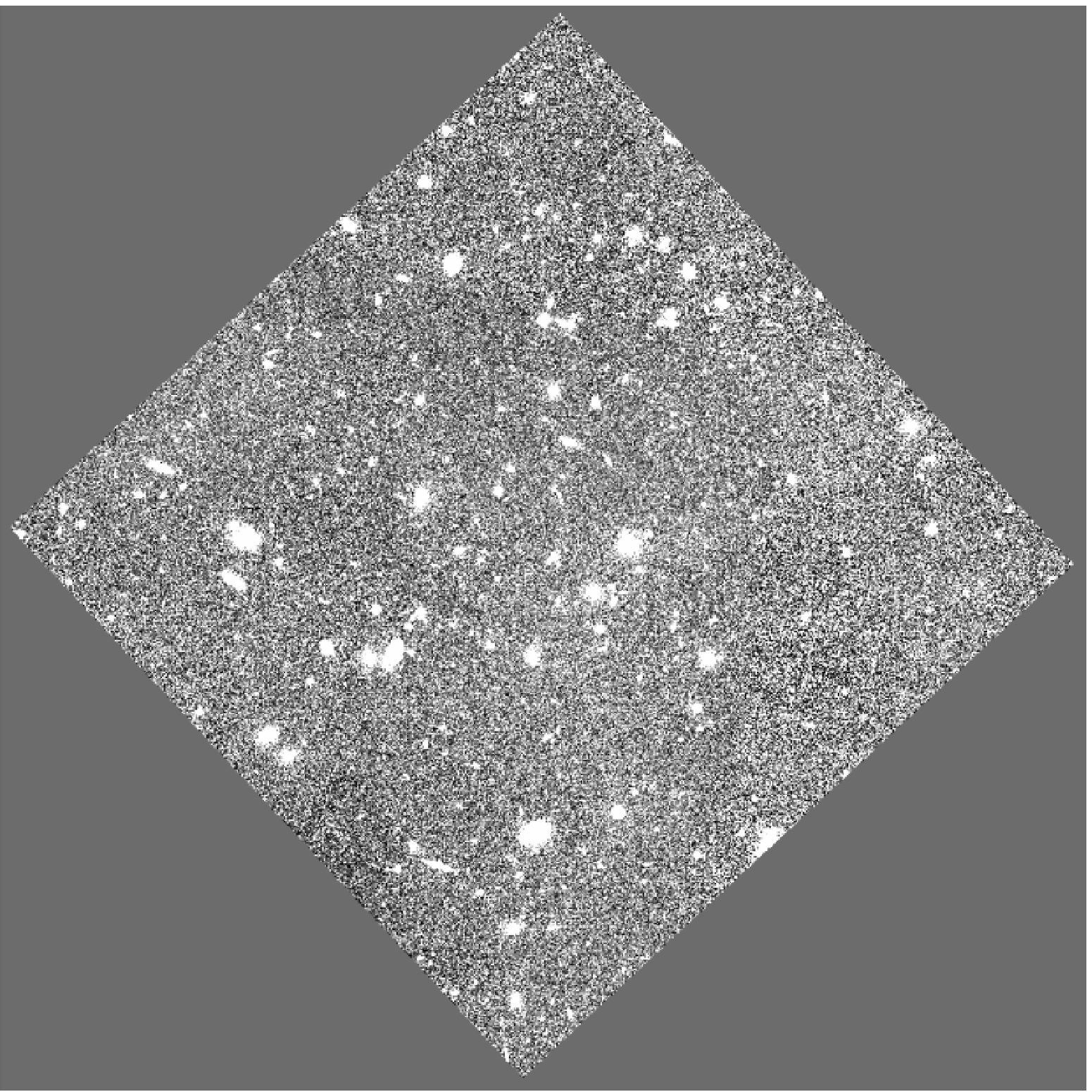}}
\centerline{\includegraphics[width=8.5cm]{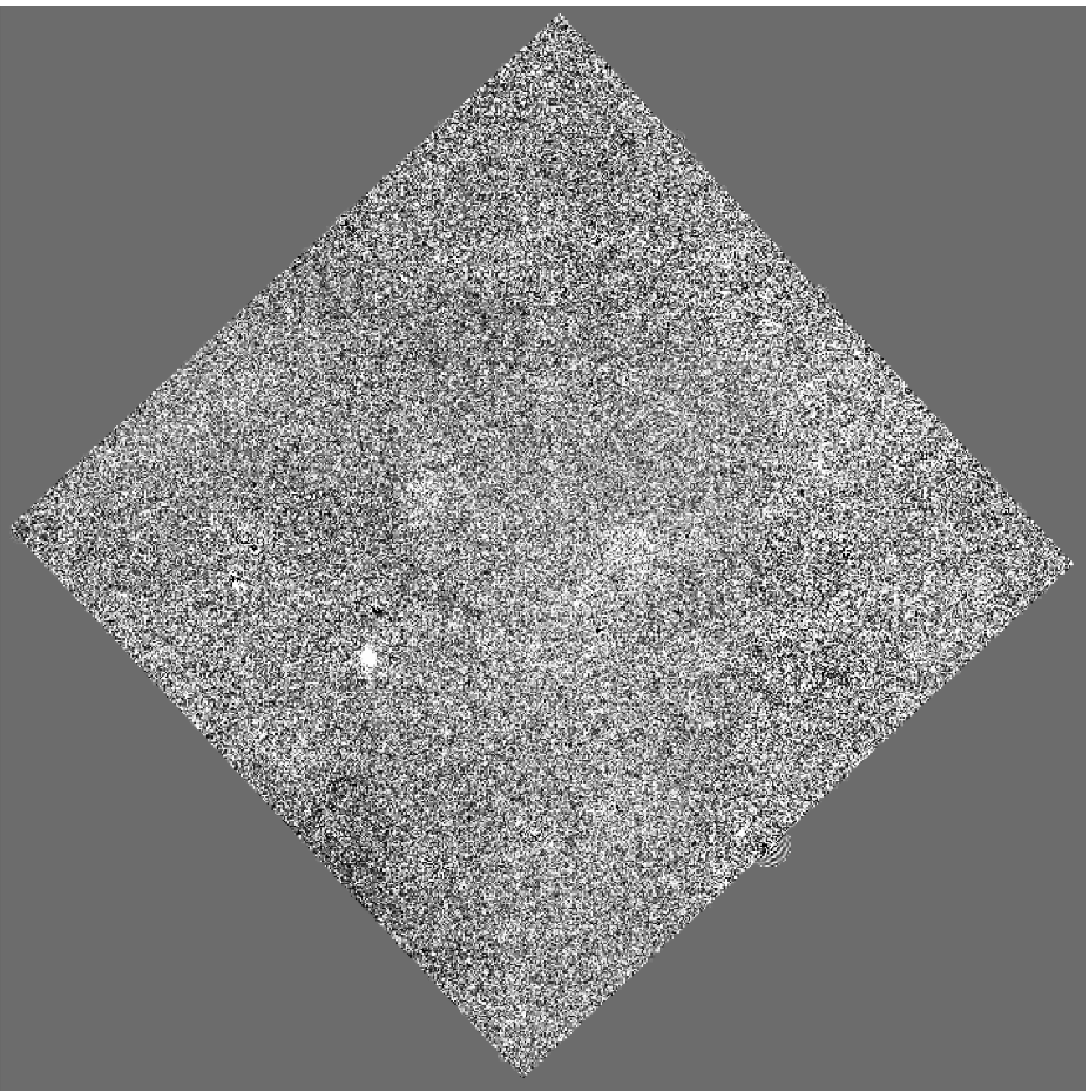}}
\caption{Original and residual images of the XDF F814W band, before and after being processed by the CHEF algorithm.} \label{814_residual}
\end{figure}

\begin{figure}
\centerline{\includegraphics[width=8.5cm]{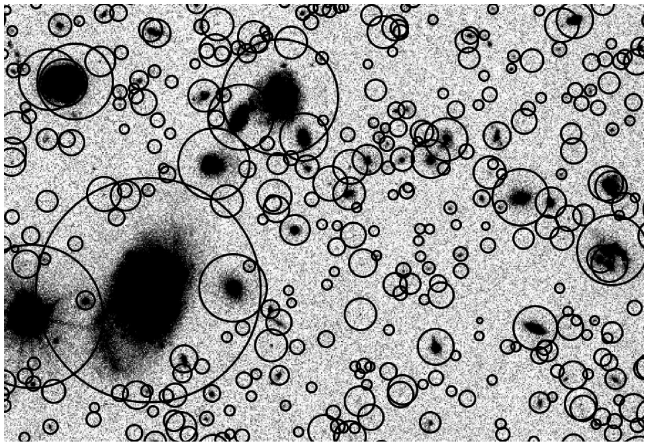}}
\caption{Frame from the XDF F814W with the apertures chosen by the CHEF algorithm.} \label{814_aperturas}
\end{figure}

In Fig. \ref{814_residual}, we display the F814W images of the original data and the residuals after the subtraction of all the CHEF models. We can observe the homogeneity of these residuals, mostly containing background and noise, and the symmetry of the remaining pixels, which ensures the photometry is not altered.  As it was explained in Section \ref{practical}, the apertures used to measure this photometry were variable, since they were settled to individually encircle all the flux from the CHEF models up to the radius where this asymptotically converges. In Figure \ref{814_aperturas} we show how these apertures vary according to the size of each source. The use of variable apertures provides an advantage over traditional aperture photometry, as does the use of the CHEF model to compute these fluxes, since we are minimizing the contamination of the noise from the fainter, outermost regions of the galaxies.\\

\subsection{BPZ settings}

 BPZ \citep{bpz,bpznew} is a SED-fitting code to accurately estimate photometric redshifts. This algorithm fits a set of optimized templates to predict the colours of the galaxies and applies a Bayesian prior to disentangle the most likely redshift solution from the final multimodal probability distribution $p(z,T)$. For this work we have used an updated version of the software, BPZ2.0, which includes several changes with respect to its original version (see \cite{molino} for more details). The new library is composed by 11 SED templates. Six of them were taken from the Projet d'\'Etude des GAlaxies par Synth\`ese \'Evolutive (PEGASE, \cite{fioc}) and re-calibrated using FIREWORKS photometry and spectroscopic redshifts
\citep{wuyts}. The other five SEDs consist on four templates from GRAphite and SILicate (GRASIL) and one STARBURST template. As for the galaxy types, the set contains five
templates for elliptical galaxies, two for spirals and four for
starbursts, along with emission lines and dust extinction. The treatment of the
opacity of the intergalactic medium was made according to \cite{madau}. BPZ2.0 also differs from the previous version in the Bayesian prior, which have been calibrated with the real redshift, spectral-type and magnitude distributions from three surveys:
GOODS-MUSIC \citep{santini}, COSMOS \citep{scoville} and  the UDF
\citep{dan}.\\
 
 To obtain precise photometric redshifts, we sample the redshift space within $0.01<z<9.0$ with intervals of DZ = 0.001. The reliability of a photometric redshift will be quantified integrating its probability distribution function in a interval of $\pm$ SIGMA\_EXPECTED=0.03, around the main peak of the redshift probability distribution function $P(z)$. This interval expresses the amount of probability comprised in the neighbourhood of the optimal value and, thus, gives an idea of how trustworthy this redshift is, in addition to the probability itself. This is what we call the $Odds$ parameter (see \cite{bpz} and \cite{molino} for a further explanation on this parameter as well as to visualize the 11 SED templates used). This reliability must not be confused with the precision of the photometric redshifts obtained, calculated through a sample of spectroscopic redshifts. \\

 In Table \ref{catalog} we show for guidance a portion of the CHEF photometric catalogue. It contains precise information on the sources detected, as for example the astrometry information, the CHEF total magnitude for each band with its respective uncertainty, the best estimation for the photometric redshift achieved by BPZ, $zb$, as well as the lower and upper limits of its confidence interval at $1\sigma$ level, $zbmin$ and $zbmax$. The catalogue also includes the most likely spectral-type for each object, $tb$, the parameter representing the reliability of the estimated photometric redshift, $Odds$,  and the resulting $\chi^2$ value of the SED-fitting algorithm.\\

\begin{table*}\centering\begin{minipage}{180mm}\caption{CHEF photometric catalogue}\label{catalog}
\begin{tabular}{c@{\hspace{2mm}}c@{\hspace{2mm}}c@{\hspace{2mm}}c@{\hspace{2mm}}c@{\hspace{2mm}}c@{\hspace{2mm}}c
@{\hspace{2mm}}c@{\hspace{2mm}}c@{\hspace{2mm}}c@{\hspace{2mm}}c@{\hspace{2mm}}c@{\hspace{2mm}}c
@{\hspace{2mm}}c@{\hspace{2mm}}c@{\hspace{2mm}}c@{\hspace{2mm}}c@{\hspace{2mm}}c}\hline
$ID$ & $RA (J2000)$ 	& $DEC (J2000)$ & $F105W$ & $eF105W$ & $F125W$ & $eF125W$ & ... & $F850LP$ & $eF850LP$ & $zb$ 	& $zbmin$ & $zbmax$ & $tb$ & $Odds$ & $\chi^2$\\
 & {\bf [deg]} & {\bf [deg]} & {\bf [mag]} & {\bf [mag]} & {\bf [mag]} & \bf{ [mag]} & ... & {\bf [mag]} & {\bf [mag]} & & & & & & &\\ \hline
1 & 53.164168 & -27.829812	& 28.665 & 0.274 & 30.016 & 0.784 & ... & 99.000 & 30.500 & 1.015 & 0.529 & 1.989 & 10.742 & 0.018 & 1.173\\
2 & 53.164897 & -27.829468	& 29.470 & 0.240 & 99.000 & 30.808 & ... & 32.868 & 13.216 & 0.560 & 0.343 & 1.227 & 10.761 & 0.100 & 1.107\\
3 & 53.164142 & -27.829253 & 99.000 & 31.117 & 99.000 & 30.808 & ... & 99.000 & 30.500 & 1.228 & 0.496 & 2.517 & 10.685 & 0.023 & 0.005\\
4 & 53.164487 & -27.829293	& 28.842 & 0.549 & 28.606 & 0.228 & ... & 32.368 & 12.058 & 1.185 & 0.765 & 1.543 & 10.459 & 0.026 & 0.221\\
5 & 53.164134 & -27.829132	& 29.488 & 0.547 & 99.000 & 30.808 & ... & 99.000 & 30.500 & 1.213 & 0.496 & 2.480 & 10.695 & 0.023 & 0.239\\
\hline
\end{tabular}
\medskip First rows of the CHEF photometric catalogue, as guideline. {\bf The catalogue includes the CHEF total magnitudes for the nine bands of the XDF, that is, F105W, F125W, F140W, F160W, F435W, F606W, F775W, F814W, and F850LP, as well as their associated errors and information related to the photometric redshifts obtained with BPZ. This catalogue is published in its entirety in the electronic edition of the journal.}
\end{minipage}\end{table*}

\subsection{Photometric redshift precision}

 \cite{dan} and \cite{lundgren} compiled a list of all publicly available spectroscopic redshifts within the HUDF, gathering 104 objects. Given that the XDF area is slightly smaller than the HUDF, just 101 sources were found to be within this field. With this sample we compared the precision reached by both the ColorPro software \citep{dan} and the CHEFs  to prove that they have a similar performance, with and without using the PSF respectively. ColorPro gets aperture-matched, PSF-corrected photometry to obtain very robust colours without degrading the quality of the whole dataset of images to that one with the worst seeing. The idea is that it defines a constant photometric aperture (for all the filters) according to the detection image, where it estimates both isophotal and AUTO  magnitudes. Then it PSF-degrades the detection image to match every individual image PSF and recalculates isophotal magnitudes. The difference between the original and "degraded" isophotal magnitudes is defined as the PSF-correction which is eventually applied to the AUTO magnitudes \citep{molino}. However, ColorPro needs to derive the PSF-model before attempting the calculation of the colours and, moreover, it is not capable of dealing with variable-PSF images. The CHEFs can measure total fluxes without needing the PSF (thus independently of its stability or variability and the photometric errors induced by that), in a fully automated and self-consistent way, simply requiring as an input the set of images to be analysed and a suitable SExtractor configuration file. \\
 
\begin{figure*}\centering
\centerline{\includegraphics[width=9cm,height=5cm]{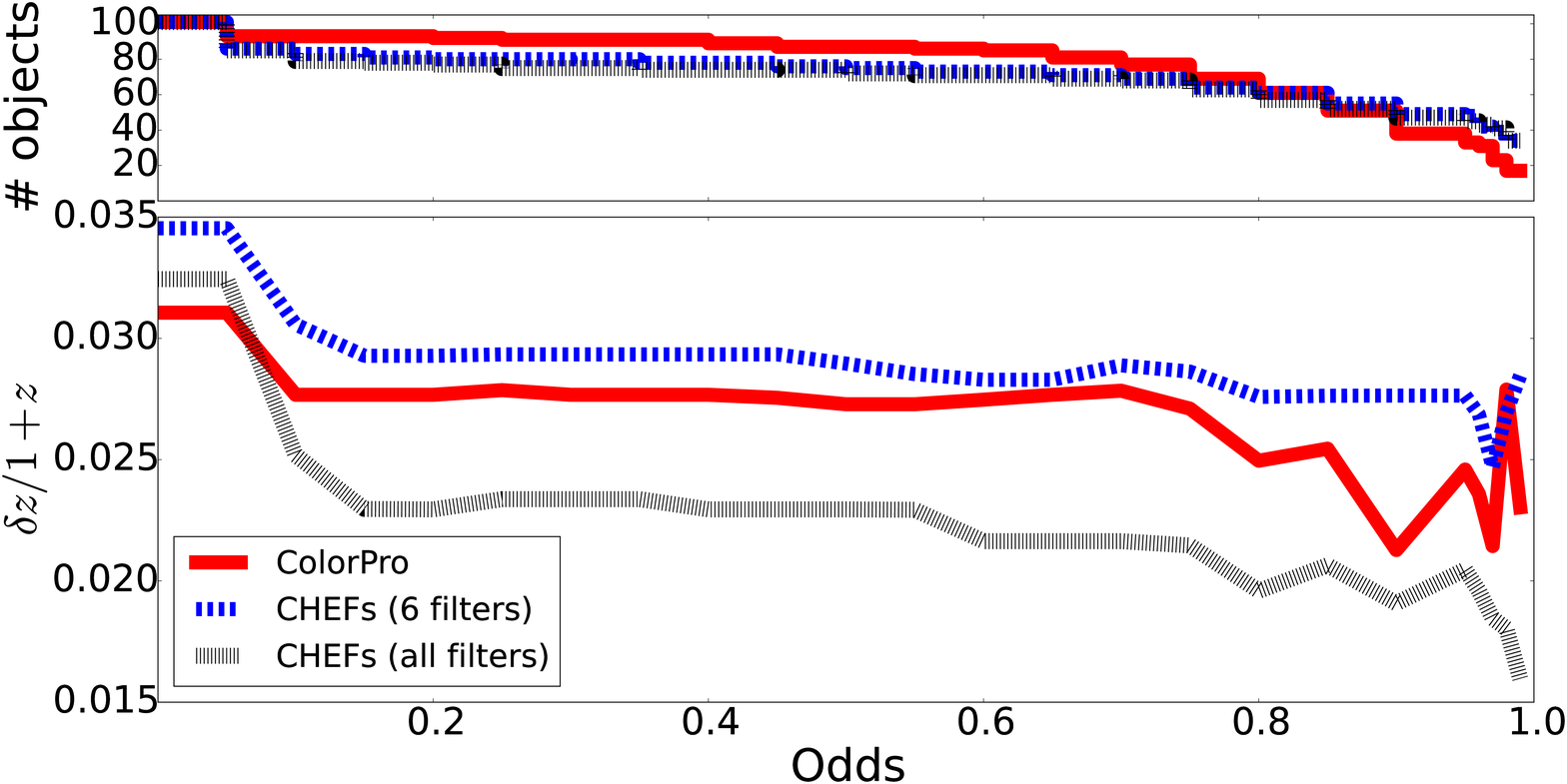}\includegraphics[width=9cm,height=5cm]{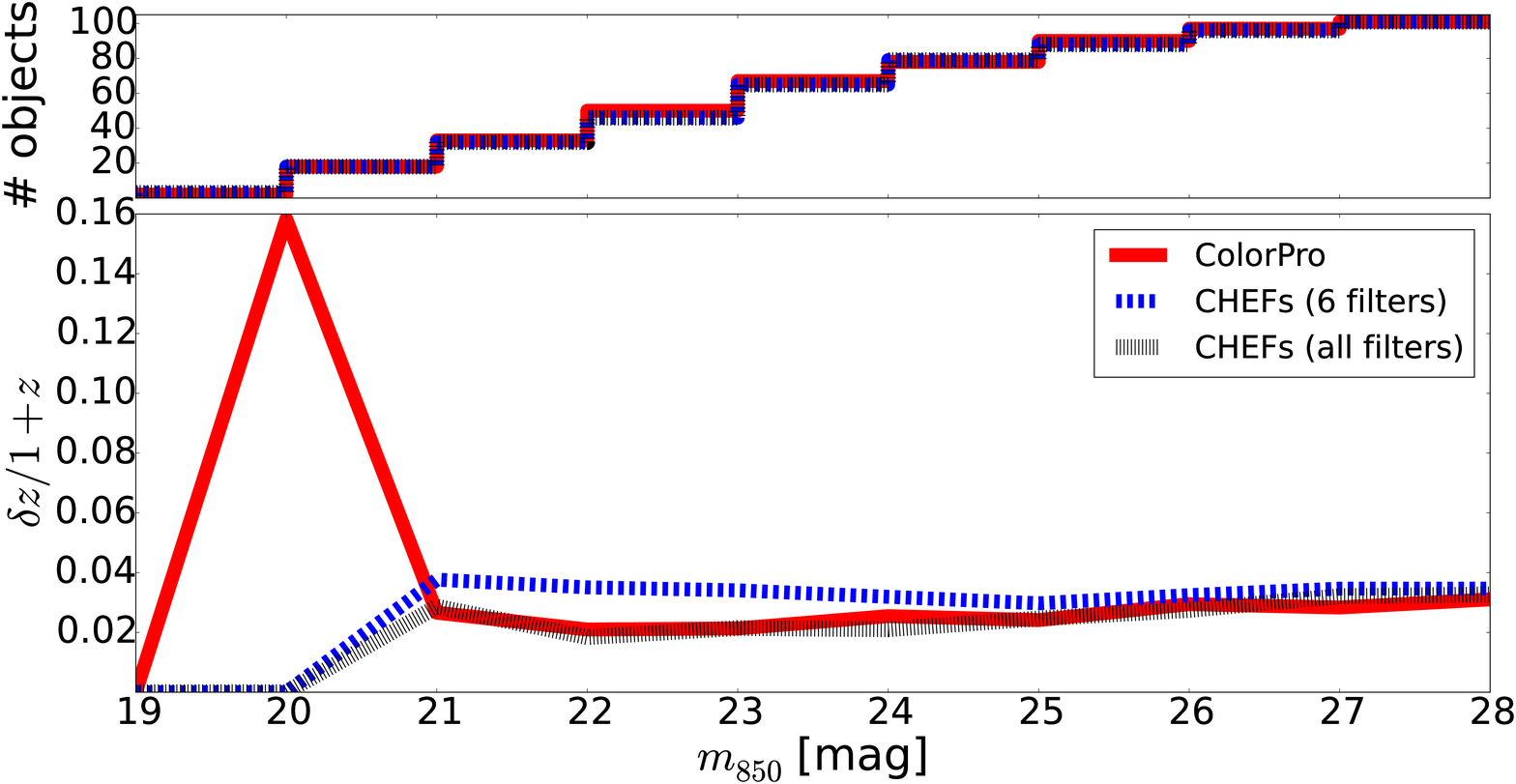}}
\caption{Comparison of the photometric redshift precisions of ColorPro and the CHEFs as a function of the $Odds$ (left) and the F850LP total magnitude (right). In both panels, the red solid lines refer to ColorPro with the HUDF data, the blue dashed lines show the precision of the CHEFs with the six XDF bands closest to the HUDF ones, and the black dotted lines display the precision of the final XDF catalogue, that is, the CHEFs performance using all the available filters. [\textit{See the electronic edition of the journal for a colour version of this figure.}]} \label{results1}
\end{figure*}
 
 The comparison of the photometric redshift results using the two codes is shown in Figure \ref{results1}. For the sake of a fair comparison between the CHEFs and ColorPro, we show not only the CHEF results using all the bands available for the XDF, but also the results taking the information from just six of the XDF filters, those most similar to the HUDF bands used in Coe's work, that is F105W, F160W, F435W, F606W, F775W, and F850LP. In this plot we show the photometric redshift precision reached by the CHEFs using the data from the XDF and the one obtained by ColorPro using the HUDF. The precision is plotted versus the $Odds$ parameter (left) and the $m_{850}$ total magnitude (right). Please notice that, for both photometries, the $Odds$ have been calculated  using the same SIGMA\_EXPECTED  value to integrate the same fraction of probability around the main peaks, so the comparison of this parameter for ColorPro and the CHEFs is consistent. When using the same number of bands, (red solid and blue dashed lines) both methods display a comparable level of accuracy as a function of the $Odds$, with a slightly better behavior for ColorPro. However, the number of objects with high $Odds$ (thus, with an unequivocal redshift estimation) is four times larger for the CHEFs than for ColorPro, which means that the CHEF colours are more stable and reliable. The CHEFs colours are better fitted by the BPZ templates, which can eventually be interpreted as a more precise (less noisy) photometry. As for the magnitudes, ColorPro behaves again slightly better than the CHEFs except for the brightest galaxies, where the CHEFs perform much more accurately. Black dotted lines show the huge improvement that the XDF data represents compared to the traditional HDUF images. The addition of the complete set of nine filters to the analysis along with the extraordinary depth of the XDF make the precision of the final CHEF catalogue increase drastically, down to 2\% of error.\\
 
\begin{figure}
\includegraphics[width=8.5cm]{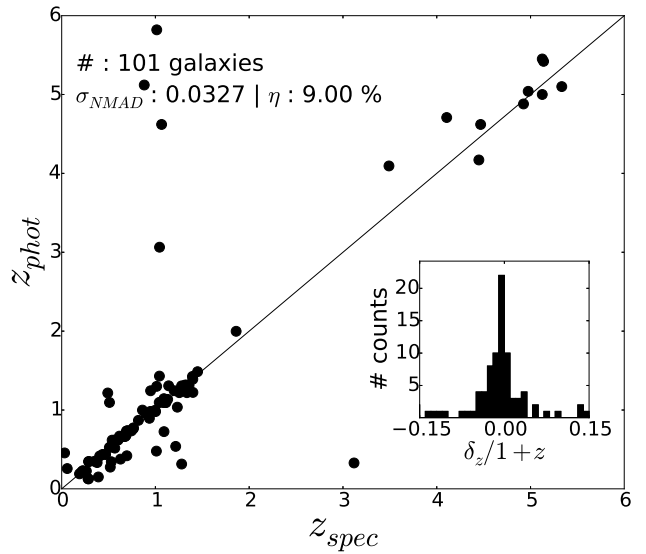}
\caption{Photometric redshift accuracy for the XDF using the CHEF photometry. We compare the photometric redshifts $z_{phot}$, obtained with the CHEFs and BPZ, to the 101 spectroscopic redshifts $z_{spec}$ available for the XDF region.} \label{results2}
\end{figure}
 
 Likewise, we compare in Figure \ref{results2} the performance of our XDF photometric redshifts (using all the available bands) against the spectroscopic sample of 101 redshifts previously mentioned. We calculate the normalized median absolute deviation, as defined by:
 
\begin{equation}
\sigma_{NMAD}=1.48\times\mbox{median }\left|\frac{\Delta z-\mbox{median }(\Delta z)}{1+z_{spec}}\right|
\end{equation}

\noindent where $\Delta z=z_{phot}-z_{spec}$. We find this scatter to be $\sigma_{NMAD}=0.0327$ for the CHEFs, value that is comparable to those measured by \cite{dan} ($\sigma_{NMAD}=0.0311$) or \cite{lundgren} ($\sigma_{NMAD}=0.055$) for the HUDF, or those obtained by \cite{bpz} (rms scatter$=0.059$) or \cite{brammer} ($\sigma_{NMAD}=0.034$) for other deep fields. We would like to remark that unlike all these works where a PSF-correction or a PSF-homogeneization has to be done for every individual image, CHEFs photometry circumvents this problem and achieves competitive or even better photometric redshift accuracy. When we compare the mean differences computed for $\Delta z/(1+z_{spec})$, we obtain better results than those found in the literature, with a mean of 0.0049 for the CHEFs versus the 0.054 got by \cite{dan} or the 0.017 yielded by \cite{lundgren}. The rate of catastrophic outliers, defined as:

\begin{equation}
\eta = \left|\frac{\Delta z}{1+z_{spec}}\right|>5\,\sigma_{NMAD},
\end{equation}

\noindent is found to be $\eta=9.0\%$, about $1\%$ lower than the obtained by \cite{dan} of $\eta=9.9\%$. Note that a small difference in the fraction of outliers will translate into a huge number of objects in large upcoming surveys like JPAS.

\subsection{XDF-UDF comparison}

\begin{figure}\centering
\includegraphics[width=8.5cm]{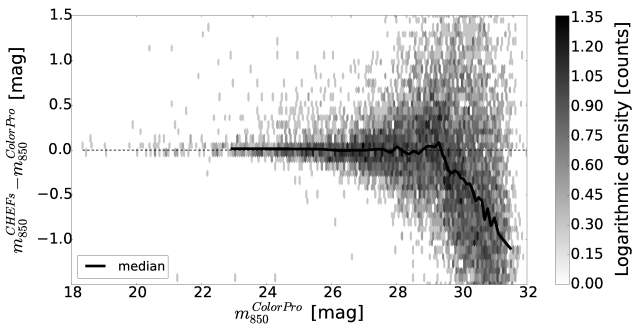}
\caption{Magnitude comparison for the XDF and UDF F850LP band in the overlapping area. A total of 14725 objects are found to be common to the two catalogues, which are plotted here in a logarithmic density plot.} \label{mag_comparison}
\end{figure}

 In addition to the comparison of the precision in the photometric redshifts obtained for the XDF and the HUDF using the CHEFs and ColorPro respectively, we also show the difference in magnitude in the overlapping area. In Figure \ref{mag_comparison} we represent a logarithmic density-map comparing both magnitudes for the F850LP band. The results show an excellent agreement for both catalogues up to magnitude $\sim$ 29.3. For fainter magnitudes, the CHEFs photometry is brighter than the ColorPro one, which can be explained by the small area used by SExtractor to compute these isophotal magnitudes (which are the base of ColorPro magnitudes, before applying the PSF correction factor).\\
 
\begin{figure}\centering
\includegraphics[width=8.5cm]{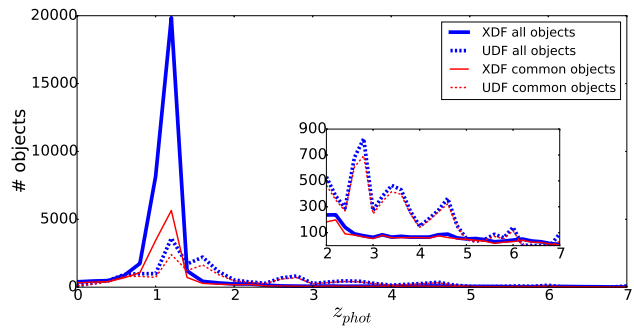}
\caption{Photometric redshift distribution for the XDF (solid lines) and the UDF (dashed lines), with the photometry calculated by the CHEFs and ColorPro, respectively. We calculate the distributions using both the complete catalogues (blue lines) and the objects in the overlapping area (red lines). The inset shows a close-up of the $z\geq 2$ region. [\textit{See the electronic edition of the
journal for a colour version of this figure.}]} \label{z_distribution}
\end{figure}

 We also study the photometric redshift distribution for the two catalogues, which can be observed in Figure \ref{z_distribution}. We represent not only the total distribution considering all the objects in each catalogue but also the distribution of the 14725 objects in the overlapping area of the XDF and the HUDF. We found an excellent agreement with the two distributions peaking at redshift 1.2 (for the complete catalogues and also for the overlapping area), although the HUDF catalogue contains a higher number of galaxies with $z\geq 2$ than the XDF.\\
 
\section{Conclusions}

 In \cite{yoli} we provided the mathematical background of the CHEF bases, and outlined some of the various applications they have beyond the modelling of the light distribution of the galaxies. In that previous work we showed that the CHEFs can be successfully used to measure the fluxes of simple, traditional profiles, such as sheared S\'ersic functions. In this paper we show the CHEFs are a powerful, precise, and highly reliable tool to calculate the photometry and the colours of real galaxies, displaying a wide variety of morphologies, brightness levels, and sizes. The CHEFs algorithm is completely automated, only needing as inputs the image to be processed and suitable SExtractor parameter configuration capable of efficiently detecting all the sources in the image. Notice that the CHEFs do not perform any detection, so they strongly rely on SExtractor results for this task. The CHEFs fit the objects extracted by SExtractor one by one, internally determining the optimal values for the CHEF parameters involved, such as the scale size $L$ or the number of Chebyshev and Fourier coefficients $n$ and $m$. Once a CHEF model is created, the algorithm uses this analytical function to accurately calculate some parameters such as the ellipticity or the real extension of the galaxy. This last quantity is particularly important for the photometry since it determines the area where the total flux of the object will be measured. With this strategy, we estimate total magnitudes, measuring them with variable apertures dependent on the size of the galaxies in the different filters and, hence, without needing the PSF. Not requiring a PSF is precisely one of the most remarkable virtue of the CHEFs: we avoid introducing new sources of error in the photometry due to a wrong determination of the PSF, inaccuracies in its centering, and variability across the image. In addition, we save a great amount of computational and human time since our algorithm is entirely automated.\\
 
  We have tested the CHEF algorithm with real optical and infrared data, both ground and space-based, affected by highly different observational conditions, background distributions and intensities, and PSFs. We have compared our results with those from the two software packages that are most widely used currently: SExtractor and ColorPro. SExtractor can calculate different types of photometry, although it does not apply any PSF correction. We have compared our total magnitudes to its MAG\_AUTO ones (using data from the HUDF and COSMOS), obtaining clearly biased results for the SExtractor measurements. We have later implemented the traditional technique of degrading all the images to the one with the worst seeing to calculate the colours using the SExtractor aperture photometry, and compared them to the CHEF colours coming from the total photometry. Again, SExtractor results remain biased while the CHEF colours are much more precise.\\
  
  We have finally tested our algorithm against ColorPro, which does not degrade the quality of the images but applies a PSF correction factor to each source. We have used the real data from the recently published XDF with the aim of not only comparing the methods with the widely used catalogue from \cite{dan}, but also creating the first public catalogue for these data. The results show that the CHEFs performance is comparable to ColorPro, with a similar level of accuracy in spite of using less information than the latter, since we do not need to estimate neither the PSF or its the effect on the photometry.\\
  
  The two catalogues strongly agree up to magnitude $\sim$ 29.3 and the redshift distributions of the sources also show a high concordance, both of them peaking at redshift $\sim$ 1.2. Our final catalogue provides precise information for 35732 objects.\\
  
\section*{Acknowledgments}

We thank the financial support provided by the CAPES program Science without Borders (project no. A062/2013) and the I3P grants from the Consejo Superior de Investigaciones Cient\'ificas.


\begin{thebibliography}{}

\bibitem[\protect\citeauthoryear{Aihara et al.}{2011}]{sdss} Aihara H. et al., 2011, ApJS, 193, 29

\bibitem[\protect\citeauthoryear{Barden et al}{2012}]{galapagos} Barden M. et al., 2012, arXiv:1203.1831v1

\bibitem[\protect\citeauthoryear{Ben\'itez}{2000}]{bpz} Ben\'itez N., 2000, ApJ, 536, 571

\bibitem[\protect\citeauthoryear{Ben\'itez}{2011}]{bpznew} Ben\'itez N., 2011, ascl: 1108.011

\bibitem[\protect\citeauthoryear{Ben\'itez et al.}{2014}]{jpas} Ben\'itez N. et al., 2014, arXiv:astroph/1403.5237

\bibitem[\protect\citeauthoryear{Bertin}{2011}]{sextractor2} Bertin E., 2001, ASP Conference Series, 442, 435

\bibitem[\protect\citeauthoryear{Bertin \& Arnouts}{1996}]{sextractor} Bertin E., Arnouts S., 1996, Astron Astrophys Sup, 117, 393

\bibitem[\protect\citeauthoryear{Bickerton \& Lupton}{2013}]{sinh} Bickerton S.J., Lupton R., 2013, MNRAS, 431, 1275

\bibitem[\protect\citeauthoryear{Brammer, van Dokkum \& Coppi}{2008}]{brammer} Brammer G.B., van Dokkum P.G., \& Coppi P., 2008, ApJ, 686, 1503

\bibitem[\protect\citeauthoryear{Boyd}{2000}]{boyd} Boyd, J.P., 2000, Chebyshev and Fourier Spectral Methods, New York

\bibitem[\protect\citeauthoryear{Coe et al.}{2006}]{dan} Coe D. et al., 2006, AJ, 132, 926

\bibitem[\protect\citeauthoryear{Fioc \&
Rocca-Volmerange}{1997}]{fioc} Fioc, M., \&
Rocca-Volmerange, B.,1997, A\& A, 326, 950

\bibitem[\protect\citeauthoryear{Illingworth et al.}{2013}]{XDF} Illingworth G.D. et al., 2013, ApJS, 209, 6

\bibitem[\protect\citeauthoryear{Jim\'enez-Teja \& Ben\'itez}{2011}]{yoli} Jim\'enez-Teja Y., Ben\'itez N., 2012, ApJ, 745, 150

\bibitem[\protect\citeauthoryear{Lundgren et al.}{2014}]{lundgren} Lundgren B.F. et al., 2014, ApJ, 780, 34

\bibitem[\protect\citeauthoryear{Lupton et al.}{1999}]{lupton} Lupton R.H., Gunn J.E., Szalay A.S., 1999, AJ, 118, 1406

\bibitem[\protect\citeauthoryear{Madau}{1995}]{madau} Madau,
P., 1995, in Meylan G., ed., QSO Absorption Lines, Springer-Verlag, Berlin, p. 377

\bibitem[\protect\citeauthoryear{Moles et al.}{2008}]{moles} Moles M. et al., 2008, AJ, 136, 1325

\bibitem[\protect\citeauthoryear{Molino et al.}{2014}]{molino} Molino et al., 2014, MNRAS, 441, 2891

\bibitem[\protect\citeauthoryear{Newberry}{1991}]{newberry} Newberry M.V., 1991, PASP, 103, 122

\bibitem[\protect\citeauthoryear{Ngan et al.}{2009}]{sersiclets} Ngan W., Van Waerbeke L., Mahdavi A., Heymans C., Hoekstra H., 2009, MNRAS, 396, 1211

\bibitem[\protect\citeauthoryear{Peng et al.}{2002}]{galfit1} Peng C.Y., Ho L.C., Impey C.D., Rix H.W., 2002, AJ, 124, 266

\bibitem[\protect\citeauthoryear{Peng et al.}{2010}]{galfit2} Peng C.Y., Ho L.C., Impey C.D., Rix H.W., 2010, AJ, 139, 2097

\bibitem[\protect\citeauthoryear{Postman et al.}{2012}]{clash} Postman et al., 2012, ApJS, 199, 25

\bibitem[\protect\citeauthoryear{Refregier}{2003}]{shapelets} Refregier A., 2003, MNRAS, 338, 35

\bibitem[\protect\citeauthoryear{Santini et
al.}{2009}]{santini} Santini, P., Fontana, A., Grazian, A., et
al., 2009, A\& A, 504, 751

\bibitem[\protect\citeauthoryear{Scoville et
al.}{2007}]{scoville} Scoville, N., Abraham, R.G., Aussel, H.,
et al., 2007, ApJS, 172, 38

\bibitem[\protect\citeauthoryear{Sesar et al.}{2011}]{sesar} Sesar B. et al., 2011, AJ, 142, 190

\bibitem[\protect\citeauthoryear{Taniguchi et al.}{2007}]{cosmos} Taniguchi Y. et al., 2007, ApJS, 172, 9

\bibitem[\protect\citeauthoryear{Wuyts et al.}{2008}]{wuyts}
Wuyts et al. 2008, ApJ, 682, 985

\end{thebibliography}
\end{document}